\documentclass[a4paper,fleqn,usenatbib]{mnras}

\usepackage{newtxtext,newtxmath}

\usepackage[T1]{fontenc}
\usepackage{ae,aecompl}

\usepackage{graphicx}	
\usepackage{amsmath}	
\usepackage{amssymb}	

\usepackage{xcolor}
\usepackage{bm}
\definecolor{mygreen}{rgb}{0.0, 0.7, 0.0}
\definecolor{mygreen2}{rgb}{0.1, 0.5, 0.1}
\definecolor{mygreen3}{rgb}{0.7, 0.0, 0.4}
\definecolor{mygreen4}{rgb}{0.1, 0.9, 0.9}
\definecolor{mygreen5}{rgb}{1.0, 0.2, 0.0}

\usepackage[figure,table]{totalcount}
\newcommand\m[1]{#1}
\newcommand\ma[1]{#1}
\newcommand\mb[1]{#1}
\newcommand\mc[1]{#1}
\newcommand\md[1]{#1}

\def\aj{AJ}%
\def\apj{ApJ}%
\def\apjl{ApJ}%
\def\apjs{ApJS}%
\def\aap{A\&A}%
\def\mnras{MNRAS}%
\def\pasp{PASP}%
\def\nat{Nature}%
\title[A connection between $\gamma$-ray and parsec-scale radio flares in the blazar 3C~273]{A connection between $\gamma$-ray and parsec-scale radio flares in the blazar 3C~273}
\author[M.~M.~Lisakov, Y.~Y.~Kovalev, T.~Savolainen, T.~Hovatta, A.M.~Kutkin]{M.~M.~Lisakov$^{1}$\thanks{E-mail:
lisakov@asc.rssi.ru}, Y.~Y.~Kovalev$^{1,2}$ , T.~Savolainen$^{2,3,4}$, T.~Hovatta$^{3,4}$, A.M.~Kutkin$^1$\\
$^{1}$Astro Space Center of Lebedev Physical Institute, Profsoyuznaya 84/32, 117997 Moscow, Russia\\
$^{2}$Max-Planck-Institut f\"{u}r Radioastronomie, Auf dem H\"{u}gel 69, 53121 Bonn, Germany\\
$^{3}$Aalto University Mets\"{a}hovi Radio Observatory, Mets\"{a}hovintie 114,02540 Kylm\"{a}l\"{a}, Finland\\
$^{4}$Aalto University Department of Radio Science and Engineering,P.O. BOX 13000, FI-00076 Aalto, Finland}

\date{Accepted 2017 March 21. Received 2017 March 21; in original form 2016 July 10}

\pubyear{2017}

\begin{document}

\label{firstpage}
\pagerange{\pageref{firstpage}--\pageref{lastpage}}
\maketitle

\begin{abstract}
We present a comprehensive \mc{5--43~GHz} VLBA study of the blazar 3C~273 \mc{initiated after an onset} of a strong $\gamma$-ray flare in this source. We have analyzed the kinematics of new-born components, light curves, and position of the apparent core to pinpoint the location of the $\gamma$-ray \mc{emission}. \mc{Estimated} location of the $\gamma$-ray emission zone is close to the jet apex, $2$~pc to $7$~pc upstream from the observed 7~mm core\mc{. This} is supported by ejection of a new component. The apparent core position was found to be inversely proportional to frequency. \m{The brightness temperature in the 7~mm core reached values up to at least $10^{13}$~K during the flare.} 
\mc{This supports the dominance of particle energy density over that of magnetic field in the 7~mm core. Particle density increased during the radio flare at the apparent jet base, affecting synchrotron opacity. This manifested itself as an apparent core shuttle along the jet during the 7~mm flare.}
It is also shown that a region \mc{where optical depth decreases from $\tau\sim1$ to $\tau\ll1$} 
spans over several parsecs along the jet. The jet bulk flow speed estimated at the level of $12c$ on the basis of time lags between \mc{7~mm} light curves of stationary jet features is 1.5 times higher than that derived from VLBI \mc{apparent} kinematics analysis.
\end{abstract}



\section{Introduction}

A connection between $\gamma$-ray activity in blazars with evolution of their parsec-scale structure and flux-density variations is now claimed by many authors \citep[e.g.][]{pushkarev_radio2gamma_delay_2010, agudo_oj287_gammasite_14pc_2011,jorstad_3c273flares} . However it is still debated from where the $\gamma$-ray emission originates\m{: on sub-parsec scales or on a few parsec scales.} The former \m{scenario} implies that \m{the} $\gamma$-ray emission zone is located close to the true beginning of the jet and the black hole \citep[e.g.][]{tavecchio_gammasite_2010,pushkarev_radio2gamma_delay_2010,rani_gammasite_2014}. Arguments for this point are the following: short timescale of variability in $\gamma$-rays implies that the emitting region should be small and cross-correlation between light curves in $\gamma$-rays and radio statistically puts $\gamma$-ray flares before radio flares. \m{$\gamma$-ray spectral breaks, if caused due to pair production, also argue in favor of the former scenario.}

The latter scenario claims that $\gamma$-ray photons should inevitably be absorbed if emitted close to the central black hole \citep[e.g.][]{ghisellini_madau_1996, ghisellini_gammasite_2008}. Moreover some $\gamma$-ray flares are reported to coincide with notable changes in flux and polarization of features located many parsecs away from the \m{jet apex} \citep[i.e.][]{agudo_0235_2011,jorstad_conf_2011,schinzel_2012}. The inconsistency of rapid variability with this scenario is usually dealt with by introducing a fine structure inside a jet with some parts moving faster or closer to the line of sight and emitting $\gamma$-rays \citep[e.g.][]{ghisellini_gammasite_2008,marscher_multizone_2014}.

If a $\gamma$-ray event occurs in a part of a relativistic jet that is optically thin for radio emission, then radio band variability is expected to be simultaneous at different radio frequencies and be simultaneous or even precede the $\gamma$-ray variability \citep[e.g.][]{ghisellini_gammasite_2008,agudo_oj287_gammasite_14pc_2011,jorstad_3c4543_gammasite_2013}.
But if a $\gamma$-ray flare arises in a region that is opaque to radio waves, then one should expect a time lag between $\gamma$-ray event and corresponding event at radio frequencies as well as a delay between activity at different radio frequencies \citep[e.g.][]{pushkarev_radio2gamma_delay_2010, fuhrmann_radiogamma_2014, ramakrishnan_radiogamma_2015}.
At the same time, one should take into account that the emitting region at radio frequencies is substantially larger than that at $\gamma$-rays. This may \m{introduce delays in the observed light curves} even if flares are ignited co-spatially and simultaneously. 
To distinguish between \m{the} two scenarios we have conducted a series of multi-frequency VLBI observations that covered a 5 month period during and after a major $\gamma$-ray flare in 3C~273 \citep{fermi_3c273}.

3C~273 is one of the most frequently observed and studied sources. It is \m{the} first discovered quasar \citep{schmidt_1963} and also \m{the} first quasar detected in $\gamma$-rays \citep{swanenburg_1978}. 
3C~273 was proven to have extremely high values of brightness temperatures in excess of $T_\mathrm{b}\approx 10^{13}$~K and features as small as 26~$\mu$as \citep{kovalev_3c273_2016, johnson_3c273_2016}. There are already several published papers on the $\gamma$-ray activity of 3C~273 \citep[e.g.][]{fermi_lc, fermi_3c273, rani_gammasite_3c273_2013}, some with single-dish radio data and $\gamma$-ray data \citep[e.g.][]{ fuhrmann_radiogamma_2014,ramakrishnan_radiogamma_2015}, and others using radio VLBI observations \citep[e.g.][]{pushkarev_radio2gamma_delay_2010, jorstad_3c273flares}.




In this paper, we aim to use the capabilities of multi-frequency VLBA observations to pinpoint the location of the $\gamma$-ray--emitting zone in the jet of 3C~273. We used measurements of the frequency-dependent apparent core position, light-curve cross-correlation, and component kinematics. 
Moreover, several by-products are discussed: a gradient of spectral index in the vicinity of the apparent jet beginning, apparent core movement, and we introduce light-curve-based method for jet speed estimation.

We adopt the $\Lambda$CDM cosmology with the following parameters: $H_\mathrm{0} = 71$~$\mathrm{km\ s^{-1}}$, $\Omega_\mathrm{m}=0.27$, $\Omega_\mathrm{\Lambda} = 0.73$ \citep{komatsu_cosmology_2009}. 
3C~273 is at a redshift of $z=0.1583$ \citep{strauss_redshift}, which corresponds to $2.7\,\mathrm{pc}\,\mathrm{mas^{-1}}$ linear scale. 

\section[]{Observations and data processing}


\subsection{Gamma-rays}

\textit{Fermi}/LAT \m{\citep[Large Area Telescope,][]{fermi_lat}} is a great facility to study temporal variability of blazars because of its sensitivity and all-sky coverage. \textit{Fermi} revealed that many of the blazars are powerful and variable in $\gamma$-rays. 

For the analysis presented here, we used \textit{Fermi}/LAT Pass~8 data to produce the 100\,MeV - 300\,GeV $\gamma$-ray light curve using the ScienceTools version v10r0p5. In the event selection we followed the LAT team recommendations for \m{the} Pass~8 data\footnote{http://fermi.gsfc.nasa.gov/ssc/data/analysis/\\documentation/Pass8\_usage.html}. We modeled a $20^\circ$ region around 3C~273 using the instrument response function P8R2\_SOURCE\_V6, Galactic diffuse model ``gll\_iem\_v06.fits'', and isotropic background model ``iso\_P8R2\_SOURCE\_V6\_v06.txt''. The light curves were binned using the adaptive binning method by \cite{lott12}, with estimated 20~$\%$ statistical flux uncertainty in each bin. The flux in each bin was then estimated using the unbinned likelihood analysis and the tool gtlike. All sources within $20^\circ$ of 3C~273 that are listed in the 3FGL catalog \citep{acero15} were included in the likelihood model. \m{In order to ensure the convergence of the fits in the weekly bins, we froze all the spectral parameters of all sources (including 3C~273) to the values reported in 3FGL. While in most cases the variability in the spectral index is modest compared to flux variations \citep{abdo_spectral_properties_2010}, we note that in bright sources, such as 3C~273, the spectral index can also vary during flares \citep[e.g.][]{rani_gammasite_3c273_2013}. Because we are not modeling the flaring in detail and only use the light curve as a guidance to activity periods in $\gamma$-rays, this will not affect any of our conclusions.}


\subsection{Radio}
After the $\gamma$-ray flux rose by a factor of 3 above its quiescent level we started our multiwavelength VLBA observations (proposal code S2087A). We conducted four \m{observations} which covered a five month long period (see Table~\ref{epochs}). The $\gamma$-ray flux rose even more throughout our observations and reached $\approx 20$ times  its average level in late September~2009 \citep{fermi_3c273}. Thus our multiwavelength observations covered different stages of the $\gamma$-ray \m{flaring activity}.

\begin{table}
\centering
\begin{tabular}{| c | c | c|}
\hline
Date & \parbox[t]{1.5cm}{\centering Number of \\antennas} & \parbox[t]{3cm}{\centering Visibilities $\times 10^3$ per band\\C, X, U, K, Q}\\ \hline
2009 Aug 28 &  10 & 5.5, 7.5, 7, 8.5, 12 \\
2009 Oct 25 &  10 & 7.5, 8, 9, 9, 15 \\
2009 Dec 05 &  10 & 5, 6.5, 6, 5.5, 12 \\
2010 Jan 26 &  9  & 6, 5.5, 6.5, 7, 10 \\
\hline
\end{tabular}
\caption{Epochs of VLBA multi-frequency observations carried out by the authors. HN antenna dropped out during the last epoch. Otherwise no major losses. Visibilities are counted with frequency and time (5~s) averaging.}
\label{epochs}
\end{table}

\begin{table}
\centering
\begin{tabular}{| c | c | c | c |}
\hline
Band & \parbox[t]{1.5cm}{\centering Wavelength\\(cm)} &\parbox[t]{1.5cm}{\centering Frequency\\(GHz)} & \parbox[t]{1.5cm}{\centering Bandwidth\\(MHz)}\\ \hline
Q1 & 0.7 & 43.2 & 32 \\ 
K1 & 1.3 & 23.8 & 32\\
U1 & 2 & 15 & 32\\
X2 & 3.6 & 8.4 & 16\\
X1 & 3.7 & 8.1 & 16\\
C2 & 6.0 & 5.0 & 16\\
C1 & 6.5 & 4.6 & 16\\
\hline
\end{tabular}
\caption{Frequency setup used in our multi-frequency VLBA observations.}
\label{freqs}
\end{table}

We observed in 5 bands (\textit{C, X, U, K, Q}), with four IF~channels per band. To equalize sensitivity and extend frequency coverage, the \textit{C} and \textit{X} bands were divided into 2 sub-bands thus resulting in 7 frequency bands (denoted further as stated in the Table~\ref{freqs} ). The data from each of these bands were then reduced individually.
Initial calibration was performed in \textsc{AIPS} \citep{aips}, imaging, self-calibration and model-fitting in \textsc{Difmap} \citep{difmap}. Following \cite{sokolovsky_coreshift_2011}, we paid special attention to \m{the} accuracy of amplitude calibration. After preliminary calibration, we performed overall gain correction per telescope/frequency/polarization/IF. We checked all available auxiliary information to find potential problems. If \m{the} gain correction exceeded 10\,$\%$, we applied it to the data. 
Then we ran self-calibration and imaging again on the gain-corrected data. This approach allowed us to estimate amplitude calibration accuracy of 10\,$\%$ in \textit{K1} and \textit{Q1} bands and 5\,$\%$ in \textit{C1,C2}, \textit{X1,X2}, \textit{U1} bands.

For flux calibration purposes, we utilized several AGN monitoring programs: 
MOJAVE at 2~cm\footnote{http://www.physics.purdue.edu/MOJAVE/allsources.html}, Boston University blazar monitoring program at 7~mm\footnote{http://www.bu.edu/blazars/VLBAproject.html}, UMRAO single dish monitoring at 6.3, 3.8 and 2~cm\footnote{http://dept.astro.lsa.umich.edu/datasets/umrao.php}, VLA calibrator survey at 6, 3.6 1.3~cm and 7~mm\footnote{http://www.aoc.nrao.edu/$\sim$smyers/calibration}.

To improve our temporal coverage at 7~mm, we used data from the Boston University blazar monitoring group. We took calibrated \textit{uv}-data and performed further analysis ourselves using the same techniques as for our data. 
The 7~mm data, used for the kinematics analysis, spanned a period of 5~years from 2008 to 2012. \m{For the brightness temperature analysis we added two more years \ma{of data} to cover the RadioAstron detection dates.} We excluded epochs with insufficient data quality (i.e. with number of antennas less than 9) from our analysis (Table~\ref{data_quality}). 

\begin{table}
\centering
\begin{tabular}{| c | c | c | c | c |}
\hline
Date & MJD & \parbox[t]{1.0cm}{\centering  $N_{\mathrm{eq}}$} & \parbox[t]{1.0cm}{\centering  Number of \\visibilities} & Comments\\ 
\hline
2008-01-17	&54482&	8	&	1829	&	-HN,  NL 60\% lost  \\ 
2008-02-29	&54525&	8	&	2433	&	-HN, -FD  \\ 
{*}2008-06-12	&54629&	6	&	1168	&	-BR, - LA, -PT  \\ 
{*}2008-07-06	&54653&	7.5	&	1956	&	-BR, -OV  \\ 
2008-08-15	&54693&	9.5	&	2408	&  \\ 
$\dotsb$ 	&$\dotsb$ &	$\dotsb$ 	&	$\dotsb$ 	&  $\dotsb$ \\ 
\hline
\end{tabular}
\centering
\caption{Epochs and data quality of 7~mm VLBA observations, both from our multifrequency campaign and from the Boston University blazar monitoring. $N_{\mathrm{eq}}$ -- roughly estimated number of antennas after accounting for all flagged data. Number of visibilities is calculated after IF averaging, integration time 20~sec. ``-ANTENNA'' in comments stands for total loss. Epochs marked with an asterisk were excluded from subsequent analysis. A full version of the table is presented in online-only materials (Table~11).}
\label{data_quality}
\end{table}

\section{Modelfitting and components parameters}

\subsection{Model fitting}

\begin{figure}
\centering
\hspace{-0.2cm}
\includegraphics[width=8cm, angle=0]{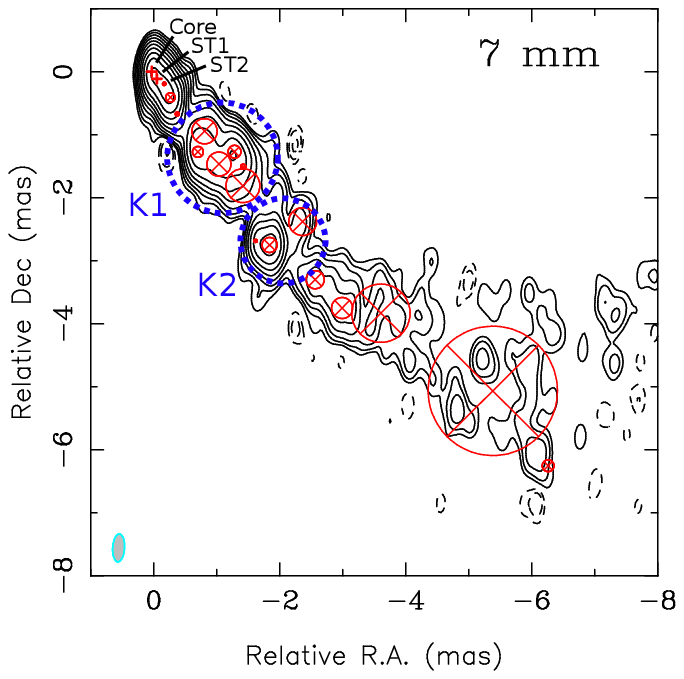}\\
\includegraphics[width=3.8cm, angle=-90]{{fig1_2}.eps}
\includegraphics[width=3.8cm, angle=-90]{{fig1_3}.eps}
\includegraphics[width=3.8cm, angle=-90]{{fig1_4}.eps}
\includegraphics[width=3.8cm, angle=-90]{{fig1_5}.eps}
\includegraphics[width=3.8cm, angle=-90]{{fig1_6}.eps}
\includegraphics[width=3.8cm, angle=-90]{{fig1_7}.eps}

\caption{Example of models obtained with model fitting for the multi-frequency VLBA observation on 2009~Aug~28. Plus signs denote $\delta$-function components while circles with crosses show the FWHM of circular Gaussian components. Axes are relative Right Ascension -- relative Declination. Units are milliarcseconds. Wavelength is denoted in the upper right corner. Beam area at half maximum is shown in lower left corner. Two dashed circles on the 7~mm map mark clusters used for analysis in Sect.~\ref{kinem:fine}. Contours are plotted with a factor of 2. Image parameters (lowest contour value [mJy~beam$^{-1}$], peak [Jy~beam$^{-1}$]) per band: 7~mm (4, 2.7), 1.3~cm (9, 5.0), 2~cm (8, 6.6), 3.6~cm (4, 6.7), 3.7~cm (3, 7.1), 6~cm (8, 11.6), 6.5~cm (11, 11.6) }
\label{modelmaps}
\end{figure}

The first step of the analysis we performed on our VLBA data was model fitting of source structure with a set of two dimensional Gaussian components (Fig.~\ref{modelmaps}). We used circular Gaussian components for \m{all} bright features of the jet structure except sometimes the core. In cases when a fitted size of a component approached zero, it was replaced by a $\delta$-function. 

3C~273 has a very broad and complex jet and the number of components in models usually exceeds 15 \m{for 7~mm and 20 for longer wavelengths} (Table~\ref{modelfit_C1}). However, cross-identification of the brightest components between epochs is \m{unambiguous and could be used} for analysis of kinematics and evolution of the components parameters. Fig.~\ref{modelmaps} illustrates models for all frequencies at a single multi-frequency epoch. Each circle represents the full width at half maximum (FWHM) of a circular Gaussian component. A full set of models is available in online materials, Fig.~16-18.

\begin{table*}
\centering
\begin{tabular}{| c | c | c | c | c | c | c | c |}
\hline
Epoch &  \parbox[t]{2.0cm}{\centering Component} &  \parbox[t]{1.5cm}{\centering Flux density \\(Jy)}    &  \parbox[t]{1.5cm}{\centering Rad\\(mas)} & \parbox[t]{1.5cm}{\centering Theta \\(deg)} & \parbox[t]{1.5cm}{\centering Major axis\\(mas)} & \parbox[t]{1.5cm}{\centering Axial ratio} & \parbox[t]{1.5cm}{\centering Phi \\(deg)}\\
\hline 
2009\_08\_28 & 0 & 6.05426 &      0.56779 &     50.40940 &            0 &      1.0 &      0.0 \\ 
         & 1 & 5.71666 &      0.64843 &   -146.68800 &      0.39853 &      1.0 &      0.0 \\ 
         & 2 & 5.30306 &      2.23402 &   -141.43400 &       1.1442 &      1.0 &      0.0 \\ 
         & 3 & 1.15141 &      3.80385 &   -134.86500 &            0 &      1.0 &      0.0 \\ 
         & 4 & 5.43772 &      5.46867 &   -126.17600 &       2.5023 &      1.0 &      0.0 \\ 
         & 5 & 0.22616 &      7.37588 &   -111.46000 &            0 &      1.0 &      0.0 \\ 
 $\dotsb$ & $\dotsb$ & $\dotsb$ &      $\dotsb$ &  $\dotsb$ &            $\dotsb$ &      $\dotsb$ &      $\dotsb$ \\ 
\hline
\end{tabular}
\caption{Modelfit parameters for all epochs. 4.6~GHz data. Component 0 represents the apparent core, other components are numbered according to their distance from the core. Rad and theta are polar coordinates of the component with respect to the map center. Major axis is the FWHM of the major axis of a Gaussian component. Axial ratio is a ratio of minor axis to major axis of a two-dimentional Gaussian component and is unity for circular ones. Phi is an inclination angle of the major axis with respect to the South--North direction. The data for all epochs and frequencies are presented in online-only materials (Tables~12--18).} 
\label{modelfit_C1}
\end{table*}

For the analysis of the brightness temperature we calculated a resolution limit for each component and used that, if the fitted size of the model component was smaller. The resolution limit was calculated following \cite{kovalev_res_limit_2005}. \ma{Noise \textit{root mean square~(rms)}, used in the signal-to-noise ratio estimation, was calculated in a \mc{$3\times3$~beam} area around the component centroid position.}


\subsection{Kinematics analysis at 7~mm}

We have performed an extensive kinematics analysis at 7~mm searching for possible correlation of flares in $\gamma$-rays with components ejection and/or flux density variations in components. 
\m{Owing to its better sampling, the shortest wavelength (7~mm) data are chosen for the jet kinematics analysis.}

We cross-identified components between epochs to trace their motion. Sometimes for a given component we had to discard several epochs from our analysis when its position became uncertain due to its low brightness or possible interaction or even merging with adjacent components. We used \m{the following} criteria to check the quality of component fitted parameters. First, position of a component should not change abruptly. 
Second, we checked that flux and brightness temperature follow a general trend to decrease with increasing separation from the core, while component size tends to increase.

As a reference point for all kinematics study we choose the VLBI core -- \m{the first component at the north-eastern} part of the jet structure. There were already published results on the presence of another stationary feature at $0.15$~mas to $0.16$~mas \citep{savolainen_3c273_1, jorstad_15agn_2005} in the 7~mm jet of 3C~273. 
Our model fitting analysis revealed two stationary features (\textit{ST1} and \textit{ST2}) besides the core \m{at a distance of} 
$r_\mathrm{ST1}=0.14 \pm 0.01$~mas and $r_\mathrm{ST2}=0.30 \pm 0.01$~mas 
from the core.
We found no components between either core and \textit{ST1} or \textit{ST1} and \textit{ST2}, hence we \m{tracked the motion of the components after passing} the most downstream stationary feature \textit{ST2}.

\m{Our model fitting results disagree with ones presented by \citet[J12]{jorstad_3c273flares}, since we used more components in our models. In particular, their core component \textit{A0} is likely to be a sum of two innermost components in our models: the core and \textit{ST1}; a short-living moving \textit{I4} from J12, which reported to be the brightest, according to light curve could be our stationary feature \textit{ST2}. This affects kinematics results and cross-identification of components with the $\gamma$-ray events.}

In \m{the} kinematic analysis, we preferred to use a linear \m{function} to fit \m{the} separation of a component from the core versus time. However there were 2 components that have shown significant deviation from a linear fit and thus were fitted with a parabola. We used a parameterization introduced by \cite{homan_kinematics_2001} to easily derive angular speed and acceleration. 

There is a known issue with parabolic fit 
that the accuracy of an extrapolation towards the core in such case would not be of a high precision. This complicates the calculation of the ejection (from the core) epoch. 

We have tested two approaches. First method is just to take \m{the first few measurements of a component separation from the core} and fit them with a straight line. Second is to use a tangent line to a parabolic fit at the position of the first data point. Both methods show consistent results within errors for both components that show accelerated motion.


\ma{Another difficulty arises when the trajectory of a component in R.A.--Dec plane does not cross the core, which is the case for the component \textit{c1}. The minimal separation does not exceed $0.1$~mas, thus we have taken the epoch of minimal separation from the core as an estimate of ejection epoch.}

\begin{figure}
\centering
\includegraphics[width=8cm]{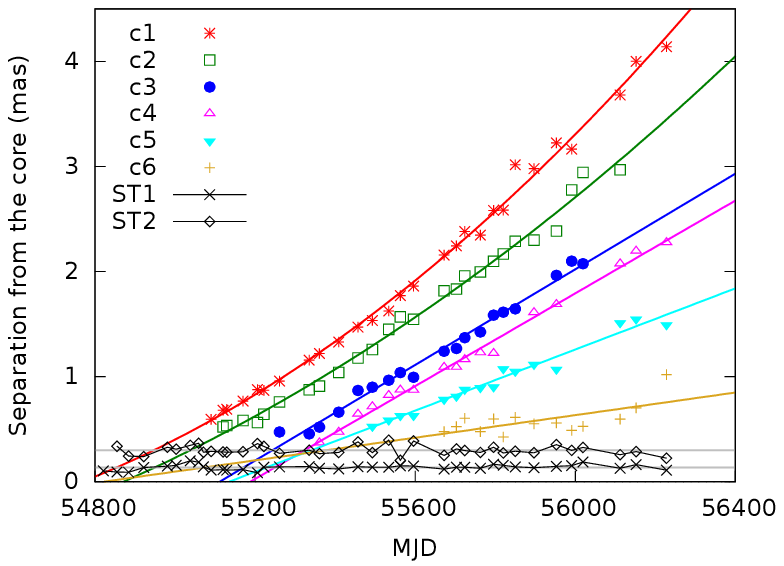}
\caption{Separation from the core for 6 components emerged in 2008--2011 and for two stationary features \textit{ST1} and \textit{ST2}.}
\label{kinem_all}
\end{figure}

We have found 6 new components during the observed period, labeled \textit{c1} to \textit{c6} according to the order of their appearance on the images. Three of them passed through the core during the period of $\gamma$-ray activity, MJD~55100 to 55300. For the last detected new component \m{\textit{c6} the fitted parameters are not well constrained, therefore we have not used it in the analysis.}
Components \textit{c1} and \textit{c2} exhibited accelerated motion, whilst for components \textit{c3}, \textit{c4}, and \textit{c5} acceleration was not significant, see Fig.~\ref{kinem_all}. Trajectories of all new components except \textit{c1} pass the core within the errors. So for the first component we used the moment of minimal separation from the core position as an estimate of the ejection epoch; for the second component we fit the innermost data points, spanning 400~days (11~measurements), with a straight line and use it for estimating the ejection epoch; for others we used simply linear fits. The derived kinematics parameters for all newborn components  are summarized in Table.~\ref{kinem}.

\begin{table*}
\centering
\begin{tabular}{| c | l | c | c | c | c |  }
\hline
Component & \parbox[t]{2.5cm}{\centering Ejection epoch\\(MJD)} & \parbox[t]{2.5cm}{\centering Radial speed\\($\mathrm{mas}\,\mathrm{yr^{-1}}$)} &\parbox[t]{3cm}{\centering Radial acceleration\\($\mathrm{mas\ yr^{-2}}$)} & \parbox[t]{3cm}{\centering max($T_\mathrm{b}$)\\(K)} \\ \hline
\textit{c1} & \qquad $54838 \pm 19$ & $1.17 \pm 0.02$ & $0.26 \pm 0.04$ & 2.2e+11 \\
\textit{c2} & \qquad $54885 \pm 43$ & $0.96 \pm 0.02$ & $0.17 \pm 0.05$ & 2.1e+11  \\
\textit{c3} & \qquad $55112 \pm 17$ & $0.83 \pm 0.02$ & -- & 4.5e+11  \\
\textit{c4} & \qquad $55189 \pm 14$ & $0.81 \pm 0.02$ & -- & 1.1e+12 \\
\textit{c5} & \qquad $55136 \pm 36$ & $0.53 \pm 0.03$ & -- & 4.6e+10 \\
\textit{c6} & \qquad $54823 \pm 381$& $0.20 \pm 0.07$ & -- & 4.1e+10 \\
\hline
\end{tabular}
\caption{Parameters of the newborn components. For components \textit{c1} and \textit{c2}, which exhibited accelerated motion, radial speed refers to the middle epoch of the  observations. 
Radial speed at the first occurrence of components \textit{c1} and \textit{c2} in the jet structure was $0.85 \pm 0.02$~$\mathrm{mas\ yr^{-1}}$ and $0.73 \pm 0.04$~$\mathrm{mas\ yr^{-1}}$ respectively. Brightness temperature is calculated in the source frame.
}
\label{kinem}
\end{table*}

\subsection{Cluster kinematics}
\label{kinem:fine}
Due to the great complexity of the jet in 3C~273, the same region of the extended structure could be model fitted differently at different epochs subject to the image sensitivity, \textit{uv}-coverage, and slight physical changes in the jet structure. This increases the  scatter of the kinematic data for each component, which is not well isolated from adjacent components. 
Therefore, to improve the accuracy of kinematic measurements, one could study kinematics of a cluster of components that describe a complex region of the jet structure, if such region is isolated from the adjacent parts of the jet.

\begin{figure}
\centering
\includegraphics[width=8cm]{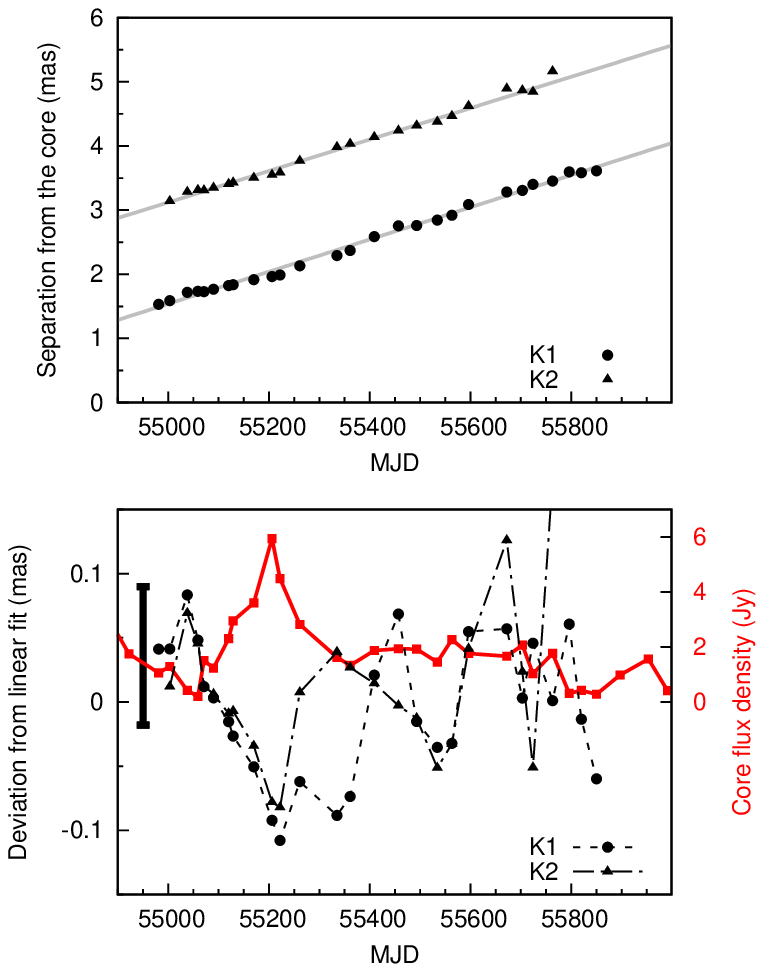}
\caption{Top: Kinematics of component clusters \textit{K1} and \textit{K2}. Bottom: Deviation from a linear fit (residuals) for \textit{K1} (circles, dashed line) and \textit{K2} (triangles, dot-dashed line) component clusters plotted together with core flux density variations(squares, red solid line). Negative values of deviation means that a cluster is closer to the core than expected from a linear fit, positive -- farther. Only a single $1\sigma$ error-bar of cluster position is plotted at the left part for better readability.}
\label{kinem_residuals}
\end{figure}

We found two well isolated clusters, \textit{K1} and \textit{K2}, in the structure of the 3C~273 jet at 7~mm during the flaring period MJD~$\approx55000-55400$, marked in Fig.~\ref{modelmaps}. 
First, we derived the position of the centroid of all components in the cluster. Coordinates were calculated as 
\begin{equation}
\bm{r} = \frac{\Sigma F_{\mathrm{i}} \bm{r_{\mathrm{i}}}}{\Sigma F_{\mathrm{i}}}
\end{equation}
where $F_{\mathrm{i}}$ is flux density, $\bm{r_{\rm i}}$ - R.A. and Dec coordinates of every single component in the cluster relative to the core at this epoch. Summation was made over all components in the cluster. 
Depending on the observing epoch, each cluster contained 2 to 6 components. We traced their position over the period and found that both clusters move with constant speed $\mu_\mathrm{K1} = 0.92 \pm 0.01$~$\rm mas~yr^{-1}$ and $\mu_\mathrm{K2}=0.89 \pm 0.02$~$\rm mas~yr^{-1}$. \m{Kinematics plot for both clusters is presented in Fig.~\ref{kinem_residuals}}. To search for fine effects we subtracted a linear fit from the kinematic data and scrutinized the residuals.

Fig.~\ref{kinem_residuals} shows the difference between the measured separation of the clusters from the core and the best linear fit for both clusters. We found that this difference for both clusters is correlated (chance probability $p=0.008$) and have contemporaneous minima. It should be noted that these clusters are $2$~mas (50~pc) away from each other, so due to causality arguments we reject the idea of their connection. This leads to the conclusion that these correlated residuals reflect not the movement of the clusters but movement of the reference point i.e. the core. 

Estimation of uncertainties of the cluster centroid position is a bit tricky. One may not use formulas suitable for isolated components because this is definitely not the case. Instead we estimated uncertainties as \textit{root mean square~(rms)} of the residuals of the linear fit to cluster kinematics. It should be noted that these estimates are upper limits, because measured residuals, as we know, are not randomly distributed.

 
%

\mb{We have detected a steady movement of the 7~mm core downstream the jet by $\approx0.17$~mas in 5~month period which corresponds to $\Delta r = 4.4$~pc along the jet for the adopted viewing angle $\theta = 6^{\circ}$.}
 

\section{CROSS-CORRELATION of light curves}


\begin{figure*}
\centering
\includegraphics[width=\textwidth]{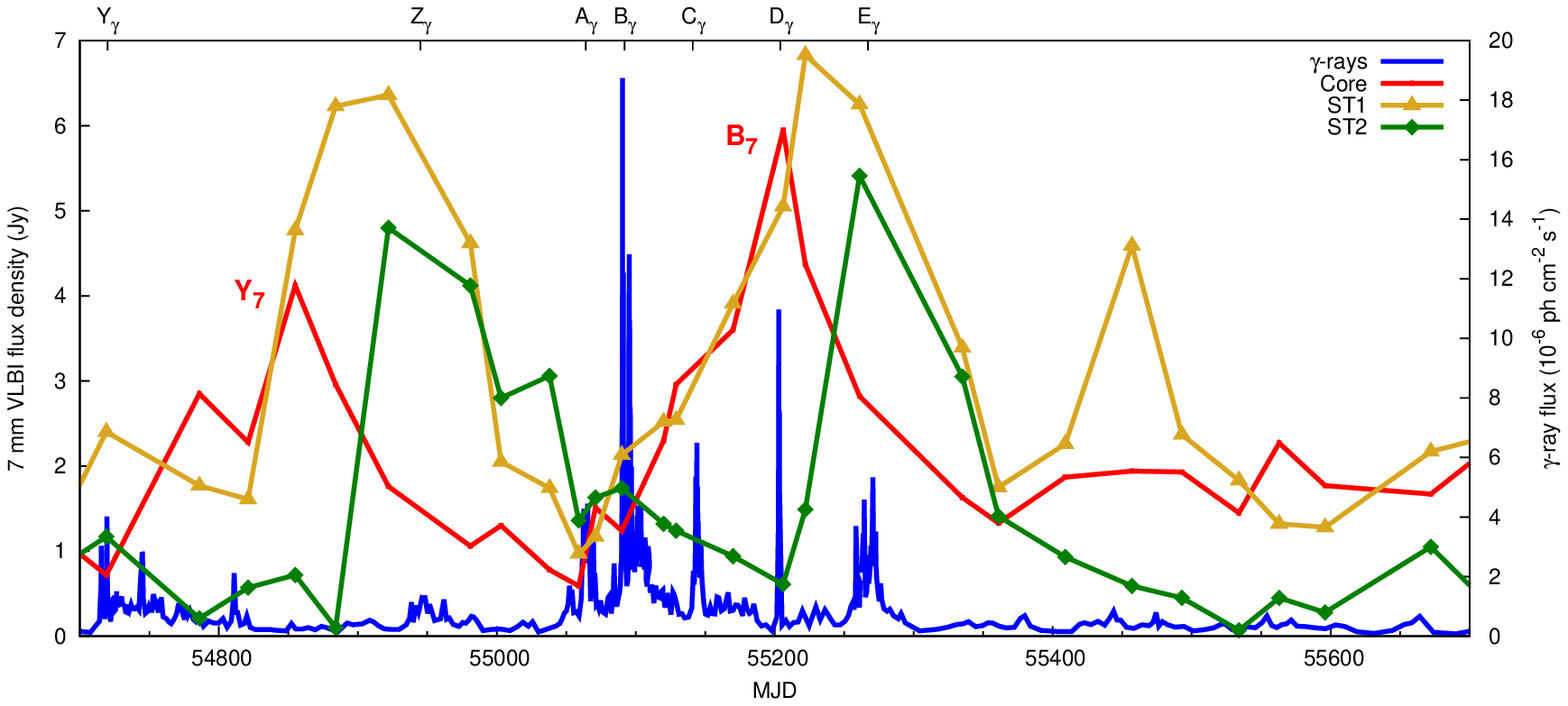}
\caption{Flux-density variation of the core at 7~mm (squares, red in online version), \textit{ST1} (triangles, amber), \textit{ST2} (diamonds, green) overlaid on the adaptively binned LAT $\gamma$-ray photon flux data (solid line, blue). $\gamma$-ray flares are denoted at the top. 7~mm core flares are labeled $Y_7$ and $B_7$.} 
\label{corestat}
\end{figure*}

Light curves of the 7~mm core and two stationary features along with the $\gamma$-ray light curve are plotted in Fig.~\ref{corestat}. Five most prominent $\gamma$-ray flares are labeled $A_{\gamma}$, $B_{\gamma}$, $C_{\gamma}$, $D_{\gamma}$, $E_{\gamma}$ following naming in \cite{rani_gammasite_3c273_2013}. Two flaring periods that occurred before $A_{\gamma}$ are labeled $Y_{\gamma}$ (several peaks around MJD~54750) and $Z_{\gamma}$ (around MJD~54950). We aimed to cross-identify flares in $\gamma$-rays and in the 7~mm radio band. Almost the whole period of $\gamma$-ray activity is covered by flares in the 7~mm radio band, which makes one-to-one cross-identification challenging. For further discussion we labeled the 7~mm core flares as $Y_7$ (peaked at MJD~54855) and $B_7$ (peaked at MJD~55206).

First, we compared the $\gamma$-ray light curve with the 7~mm light curves of the core and both stationary components.
These components are usually the brightest among all components of 3C~273 at this wavelength. Each of these components underwent 2 major flares with peak flux densities of $4$ to $7$~Jy. It is worth to note that none of these 7~mm radio flares coincided with the major flare $B_{\gamma}$. The only two peak-to-peak coincidences were: $B_{7}$ in the core coincided with $D_{\gamma}$, and second flare in \textit{ST2} ($MJD=55261$) -- with $E_{\gamma}$. 

Second, we ensured that no other moving component except the newborn components shows any notable flares during the period analyzed. Generally the flux density of all moving components tended to decrease monotonically with distance. Only the brightest ones exhibited flux density of an order of 1~Jy.

Third, we analyzed light curves of new components \textit{c1} to \textit{c6}.  Component \textit{c1} had its flux density maximum coinciding with the $\gamma$-ray flare $B_{\gamma}$. However, this maximum was the first data point for this component after separation from \textit{ST2} (it was already $0.6$~mas from the core and was $0.15$~mas across) and its flux density was $\approx2.5$~Jy, lower than peaks of the core or stationary features.

\begin{figure}
\centering
\includegraphics[width=8cm]{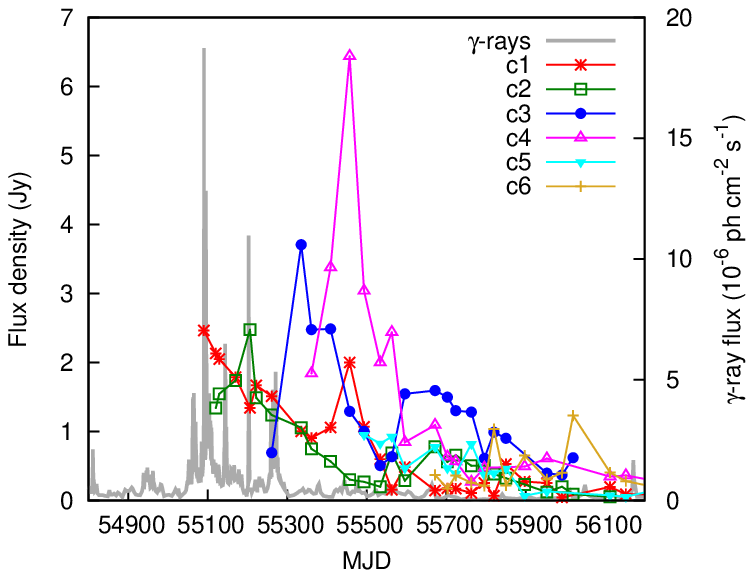}
\caption{Flux density of all new components \mc{at 7~mm} emerged in 2008--2011 overlaid on the LAT $\gamma$-ray light curve.}
\label{flux_all}
\end{figure}

\subsection{Gamma-ray to 7~mm core time lag}

\begin{figure}
\includegraphics[width=8cm]{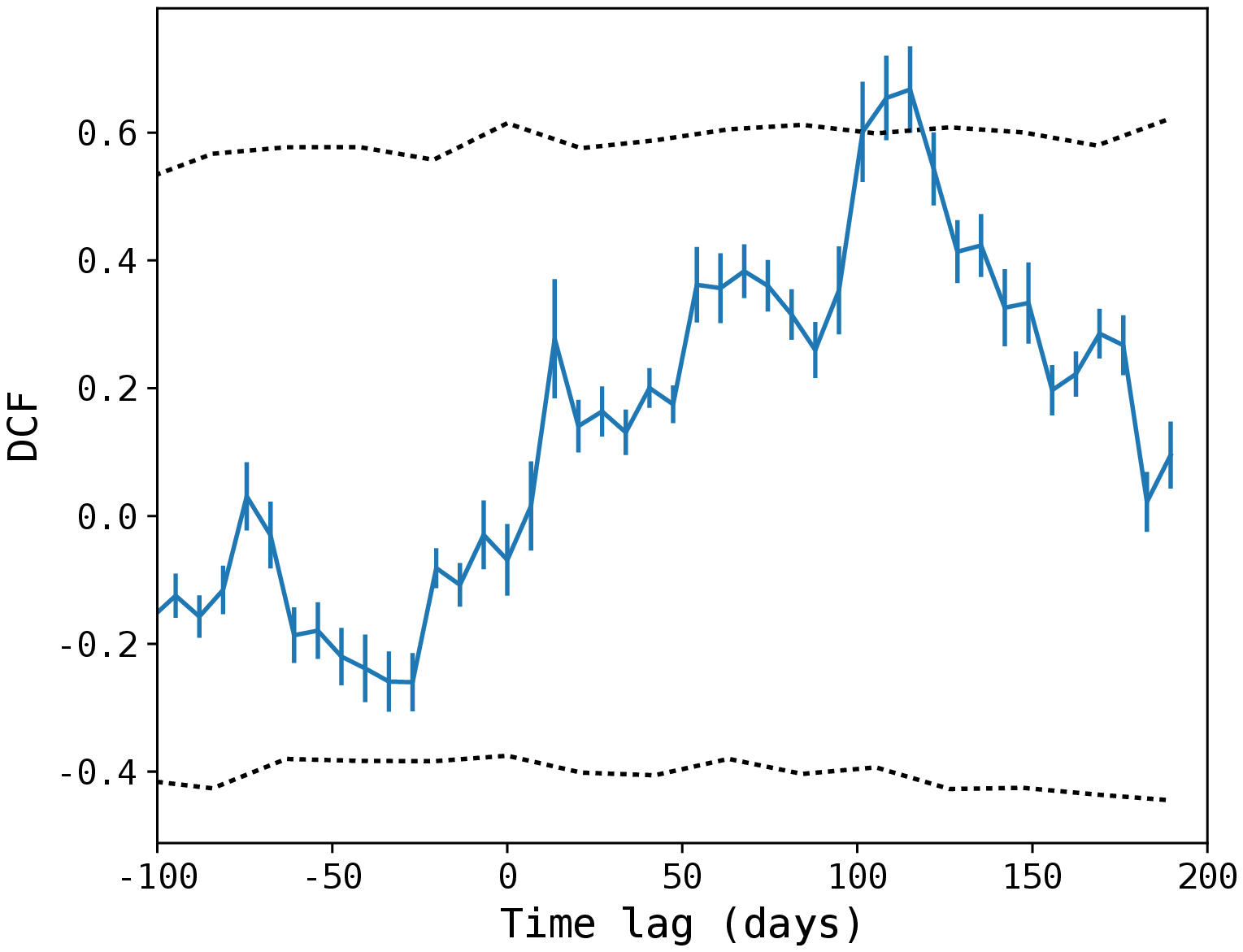}
\caption{Discrete correlation function of $\gamma$-ray to 7~mm radio core light curves (solid curve) and \m{95\,\% confidence levels} for positive and negative correlation coefficients (dotted curves).}
\label{dcf}
\end{figure}

We have analyzed the cross-correlation of the $\gamma$-ray and 7~mm radio core light curves using \m{the} discrete correlation function \citep[DCF,][]{edelson_krolik}. \m{DCF analysis curves are presented} in Fig.~\ref{dcf}.

The statistical significance of the correlation was estimated by simulation of artificial light curves. In case of $\gamma$-ray data we created a set of simulated light curves with the same power spectral density (PSD) and probability distribution functions (PDF) as the original data \citep{Emmanoulopoulos_2013, Connolly_2015}. 

The 7~mm data contain too few measurements to obtain good PSD and PDF distributions so we applied another approach in this case. For the 7~mm data, each simulated light curve was made from the original data as follows: in an array ($t_i$, $S_i$) of time and flux density  we took a random number $k$~($1 < k \leq N$, where $N$ is the total number of 7~mm data points) of flux density measurements from the end of the array and placed them at the beginning, thus transforming initial array $(t_1,t_2\ldots t_N; S_1,S_2\ldots S_N)$ into $(t_1,t_2\ldots t_N; S_k, S_{k+1}\ldots S_N,S_1,S_2\ldots S_{k-1})$.

We performed 1000 simulations calculating DCF between an artificial $\gamma$-ray and 7~mm core light curves. This resulted in a distribution of 1000 correlation functions of two unrelated signals. We then calculated a \m{$95\,\%$ confidence level} as a value such that a probability of DCF to be higher by chance than this value is $5\,\%$. Such a one-tailed test can also be applied to the negative correlation coefficients (note its absolute value is much lower due to the properties of the PDF distribution of the $\gamma$-ray data).

\begin{figure}
\includegraphics[width=8cm]{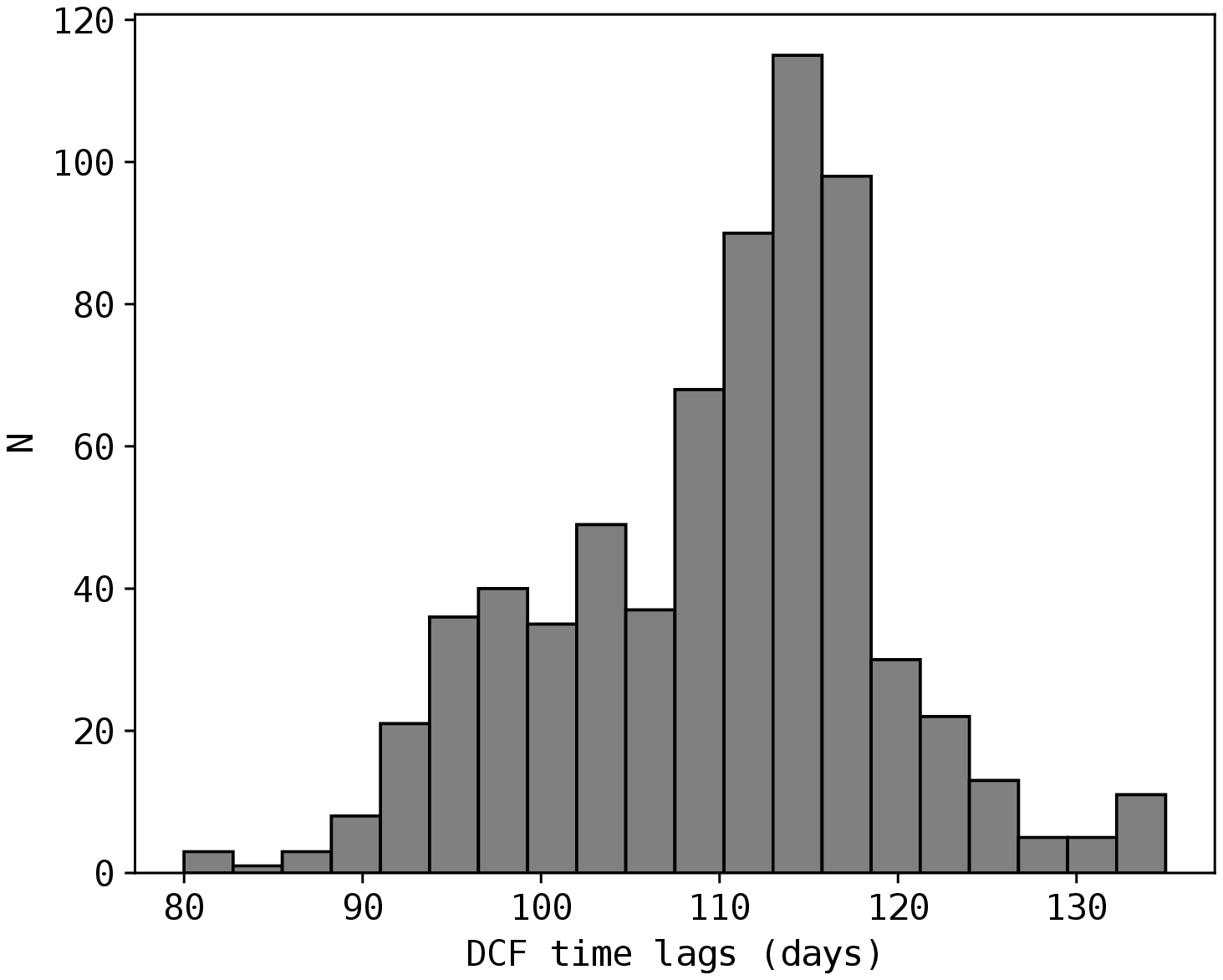}
\caption{Cross-correlation peak distribution obtained with 1000 FR/RSS simulations for light curve lags in the range  80--140~days.}
\label{ccpd}
\end{figure}

Based on these tests we find that the correlation between $\gamma$-ray and 7~mm light curves is significant for a time lag in a range 90 to 130~days. A more accurate estimate of uncertainties could be obtained using this information along with Monte Carlo simulations of the data. Now we perform 1000 evaluations of DCF using flux randomization (FR) and random subsample selection (RSS) methods \citep{peterson}. First, at each stage the flux density of every point is taken as a random value normally distributed around the measured value with dispersion corresponding to its error (FR). Second, a random subsample of the same length as the data is selected from the original light curve (RSS -- modified bootstrap). The former step (FR) accounts for uncertainties due to flux measurements, while the latter (RSS) allows one to eliminate the contribution of outliers to the correlation results. Finally we obtain a cross-correlation peak distribution (Fig.~\ref{ccpd}) for time lags in the range $80-140$~days, from which the time lag and its error were estimated as the mean and standard deviation. We obtain $\Delta t_{\mathrm{7~mm - \gamma}}^{\mathrm{obs}} = 112 \pm 9$~days lag with $\gamma$-rays leading the 7~mm  core. This value is close to the peak-to-peak lag of $115$~days between $B_{\gamma}$ and $B_7$.

For our calculations we used a DCF time bin $\delta t = 21$ days, which is the mean time spacing of $\gamma$-ray and radio light curves. We note, however, that the correlation results do not exhibit any dependence on the choice of the $\delta t$ value in the range $5\leq \delta t \leq 50$. 


We compared the 7~mm to $\gamma$-ray lag with that derived with single-dish observations at 8~mm by \cite{ramakrishnan_radiogamma_2015} who found a lag of 160~days \m{and at 7~mm by \cite{chidiac_3c273_2016} who found a lag of 160~days}. These values agree well with our 112~day $\gamma$-ray to 7~mm core lag, considering that the single-dish observations could not disentangle flares of comparable amplitudes in the core and stationary components. We checked this by estimating the time lag from the peak of the $\gamma$-ray flare $B_{\mathrm{\gamma}}$ to the peak of the aggregate core+\textit{ST1}+\textit{ST2} light curve and found a lag of 170~days.

\subsection{Subsequent flares in core--\textit{ST1}--\textit{ST2}}
\label{corest12}
Fig.~\ref{corestat} shows the light curves of $\gamma$-rays and the core and two stationary components in the 7~mm band. It is clear that a radio flare first occurs in the core and then subsequently in the two stationary components. Taking components' proper motion to be $\mu=0.8$~$\mathrm{mas\, yr^{-1}}$ -- typical for newborn components at that period -- and taking the measured distances between stationary features, one would expect the peak of the  ST2 light curve to occur at MJD~$~55400$, significantly later than observed. 

This inspired us to calculate speed of a moving disturbance that could produce such flares. Cross-correlation of light curves is inappropriate in cases of such scarcely sampled and small data samples. So to get the \m{moment $t_{\mathrm{peak}}$ of the peak} we fitted flares with an exponential rise and decay 
\begin{equation}
 S(t) =
  \begin{cases}
     S_{\mathrm{max}}\ e^{(t-t_{\mathrm{peak}})/(\tau_{\uparrow})}  ,     & \quad t<t_{\mathrm{peak}}, \\
     S_{\mathrm{max}}\ e^{(t_{\mathrm{peak}}-t)/(\tau_{\downarrow})} ,      & \quad t>t_{\mathrm{peak}},\\
  \end{cases}
\label{exp_flares}
\end{equation}
where \mc{$\tau_{\uparrow}$ and $\tau_{\downarrow}$ are variability time scales for the rise and the decay of a flare}, \m{$S_{\mathrm{max}}$ is the peak flux density}, \md{similarly to \cite{valtaoja_1999}, but with no assumed relation between $\tau_{\uparrow}$ and $\tau_{\downarrow}$}.

This worked well for the core and \textit{ST1}, but \textit{ST2} has even worse sampling, so we chose the position of a maximum in the middle between 2 points with the highest flux density with uncertainty as large as to let the true maximum fall between adjacent points with $95\,\%$ probability. 

This resulted in a proper motion $\mu = (1.21 \pm 0.06)$~$\mathrm{mas\, yr^{-1}}$ or $\beta = (12.3 \pm 0.6)\, c$ which is not extraordinary for 3C~273 \citep[e.g.][]{savolainen_3c273_1, mojave_vi_kinem, jorstad_3c273flares} but is 1.5 times higher than the speed measured downstream for the components that may be associated with these flares. 

This approach is important in two senses. First, this is an alternative way to estimate the jet plasma flow speed. Second, speed is measured very close to the apparent jet beginning where direct component kinematics usually gives poor quality due to the strong dependence between adjacent components parameters. 


\m{The difference in speed \mb{estimated near the apparent core position using the flux density light curves of the 7~mm core and stationary features and that calculated using direct position measurements farther downstream} might be explained by either different jet geometry in the core region, or some kind of a deceleration, or the difference between the flow speed and the pattern speed.}


\subsection{Brightness temperature of the core and ST1 components}
\label{britem}
\mb{With measured flux density and size of an elliptical Gaussian component one could estimate its brightness temperature in \mc{the rest frame of the source} as follows \citep[e.g.,][]{kovalev_res_limit_2005}:}
\begin{equation}
T_\mathrm{b} = \frac{2 \ln{2}}{\pi k_\mathrm{B}}  \frac{(1+z) S \lambda^2 }{\theta_\mathrm{min} \theta_\mathrm{maj}},
\end{equation}
\mb{where $k_\mathrm{B}$ is the Boltzmann constant, $\lambda$ is the wavelength, $S$ is flux density of the component, and $\theta_\mathrm{min}$ and $\theta_\mathrm{maj}$ represent either the measured FWHMs of the component or the resolution limits along the component's axes if the component is not resolved.}

\begin{figure*}
\includegraphics[width=16cm]{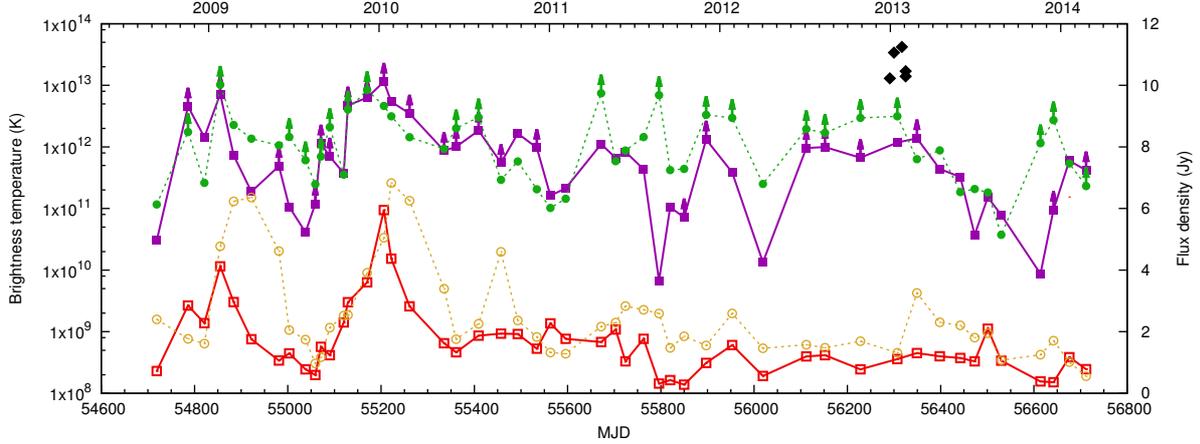}
\caption{Brightness temperature in logarithmic scale for the 7~mm core (solid purple line, filled squares), 7~mm ST1 component (dashed green line, filled circles), and for the RadioAstron detections at 18, 6, and 1.35~cm (solid black diamonds). The arrows indicate lower limits on the brightness temperature. Flux density is plotted for comparison: the 7~mm core flux density (solid red line, open squares), 7~mm ST1 flux density (amber dashed line, open circles).}
\label{tb}
\end{figure*}
\m{We found that the brightness temperature for the most compact and prominent features, the core and ST1, in the parsec-scale structure of 3C~273 could vary by several orders of magnitude, from $10^{10}$~K to $10^{13}$~K in the source frame. Throughout our multi-frequency observations the brightness temperature was of the order of several times $10^{12}$~K at all observed frequencies.}

\m{We used the long-term monitoring data at 7~mm to trace the evolution of the brightness temperature on the timescale of years. Figure~\ref{tb} shows the evolution of the brightness temperature for the most prominent features at the apparent beginning of the jet at 7~mm, the core and ST1. Since both these components were unresolved at some epochs, we used the upper limits on their size for those epochs, following \cite{kovalev_res_limit_2005}, to estimate the lower limit on the brightness temperature.}

\mb{The observed brightness temperature of the core \mc{in the source frame} varies from $10^{10}$~K to $>10^{13}$~K (Fig.~\ref{tb}), reaching the highest values simultaneously with the peak of the flare $B_7$ \mc{on 2010~Jan~10}. The highest measured values and lower limits agree with that derived by  RadioAstron \citep{kovalev_3c273_2016,johnson_3c273_2016}.
Meanwhile, the \mc{Doppler-boosted} equipartition brightness temperature \citep[][eq.\,4b]{readhead_tb_1994} \mc{in the rest frame of} 3C~273 is $T'_\mathrm{eq} \approx 4\times10^{11}$~K, for Doppler factor $\delta = 6$ \citep{savolainen_3c273_1}. According to, \mc{e.g.,}~\cite{readhead_tb_1994} the excess of a brightness temperature over the equipartition value could be ascribed to the dominance of the particle energy density over that of the magnetic field. Measured values of the brightness temperature of the order of $10^{13}$~K require that the conditions at the apparent 7~mm core should significantly deviate from equipartition (see also \cite{kovalev_3c273_2016}). During the flare the particle energy density should be greater than the magnetic field energy density.
We note, that the huge range of the $T_\mathrm{b}$ variations (Fig.~\ref{tb}) could not be explained by the variations of the Doppler factor and is most probably due to variations in the particle density and the magnetic field strength.}

\section{Frequency-dependent apparent core position}
\label{coreshift}

We have taken advantage of our multi-frequency dataset and performed analysis of the frequency dependent apparent core position, the so-called ``core~shift'' effect \citep[e.g.][]{marcaide_coreshift_1984,lobanov_coreshift_1998, kovalev_opacity_2008, osullivan_2009,  sokolovsky_coreshift_2011,kravchenko_1030_2016}. 
At each observed frequency the core is believed to be a surface of optical depth $\tau=1$ and its location \m{depends} on frequency  \citep[e.g., inversely, in model of][]{blandford_koenigl}. 

With the assumption that optically-thin synchrotron radio emission from the jet originates in the same volume at different frequencies, the position of such optically-thin features should not be affected by opacity effects and thus could be used as a reference point to derive a frequency dependent shift of optically thick parts of the jet. 
\m{We used two methods to derive the shifts between maps: a cross-correlation of large optically-thin parts of the jet structure and alignment of distinct model components \citep[e.g.][]{mojave_ix_opacity_2012, kutkin_3c454_2014}.}

In the former method a shift between two maps is derived at first using \ma{two dimensional cross-correlation} (2DCC) on the optically-thin portions of the jet structure. Then a core shift is calculated using the map shift and core coordinates in each map. The latter method uses distinct optically-thin model components at different frequencies as reference points because they should not experience a frequency-dependent shift. A component used in this method ought to be optically thin and preferably isolated to avoid contamination with adjacent components.
\m{With our model fitting strategy, all strong components in the jet structure are usually not isolated and hence their position is affected by adjacent components.}
Therefore, we just checked that optically-thin component alignment gave results consistent with the 2DCC method, but for our subsequent analysis we used only results obtained with the 2DCC method.  


For 2DCC analysis we used maps with identical map size, pixel size, beam parameters, and range of sampled baselines (\textit{uv}-range). Therefore we had to cut off data from long baselines at higher frequency and data from short baselines at lower frequency to match the \textit{uv}-range.  
\mb{We cross-correlated all pairs of maps that exhibited sufficient source structure, not only at adjacent frequencies. Then for each pair of frequencies we calculated the avareage map shift $\langle\Delta r_\mathrm{map}\rangle$ and its \textit{rms} $\sigma_{\Delta r_\mathrm{map}}$.}


We inspected spectral-index maps with the map shift applied in order to check the fidelity of the derived shift. In all cases the resulting spectral-index map looked smoother and exhibited an increased spectral index $\alpha$ ($S\propto\nu^{\alpha}$) near the core position indicating that the jet is optically thick in this region. 

\mb{The core shift comprises the map shift $\langle\Delta r_\mathrm{map}\rangle$ and the position of the cores.} The total error of the core shift derived using 2DCC consists of pixel size, core position error and $\sigma_{\Delta r_\mathrm{map}}$. To decrease the impact of pixel size on the resulting errors in core shift we reduced the pixel size to values small as $0.01$~mas which is about $1/20$th of the beam FWHM at 7~mm and a much lower value for longer wavelengths.
Areas on the maps for performing the cross-correlation were chosen manually, so we ran 2DCC several times for each frequency pair per epoch to reduce the effect of a human-driven choice. Errors in the position of the core, $\sigma_{\mathrm{core}}$, were estimated \md{in the image plane} following \cite{fomalont_errors}.

\subsection{Derived physical parameters of the jet}
\begin{figure*}
\includegraphics[width=5cm]{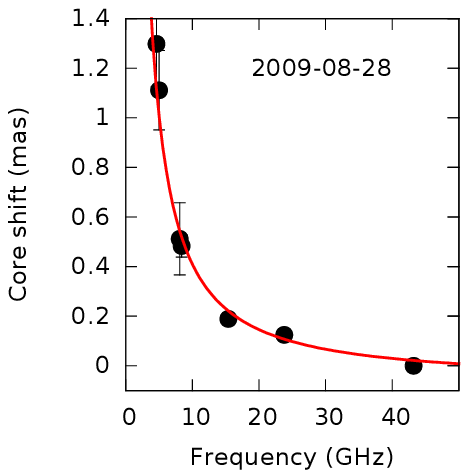}
\hspace{-15mm}
\includegraphics[width=5cm]{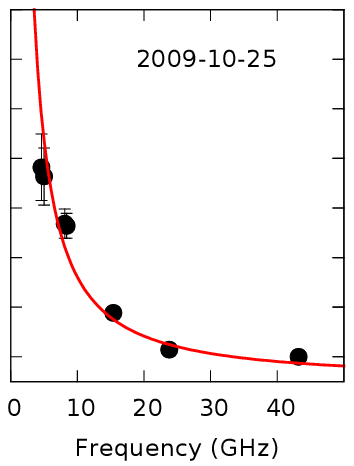}
\hspace{-15mm}
\includegraphics[width=5cm]{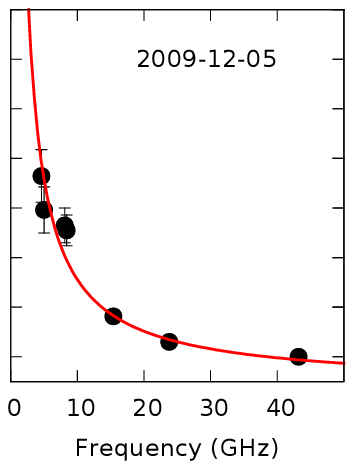}
\hspace{-5.5mm}
\includegraphics[width=5cm]{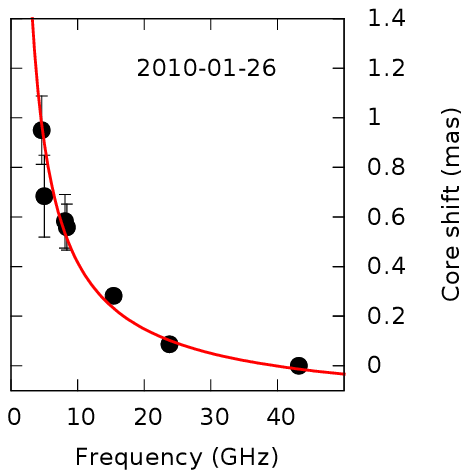}
\caption{Shift of the apparent core position versus frequency for all multi-frequency observational epochs. Solid line represents the best power-law fit.}
\label{k_fit}
\end{figure*}

%
We fitted our core-shift measurements with a power law  $\Delta r = a\nu^{-\frac{1}{k_\mathrm{r}}} +b$ as shown in Fig.~\ref{k_fit}. A model of the jet emission \citep{blandford_koenigl} assuming equipartition between magnetic field and kinetic energy and a conical shape of the jet implies $k_\mathrm{r} \approx 1$  \citep{lobanov_coreshift_1998}. 
The value of $k_r = ((3-2\alpha)m+2n-2)/(5-2\alpha)$ depends on the optically thin spectral index $\alpha$ and power indices of the B-field and particle density decay along the jet, $B(r)=B_1 r^{-m}$  and $N(r)=N_1 r^{-n}$ respectively, where $B_1$ and $N_1$ are measured at 1~pc from the jet beginning.

\mb{In the model of \cite{blandford_koenigl} the apparent jet base is located farther downstream at lower frequencies due to \mc{frequency dependent} synchrotron opacity. Thus, if a flare is caused by a disturbance that moves downstream and carries increased values of the particle density or magnetic field, then our single-epoch observation \mc{could} catch enhanced values of $N$ and $B$ in the apparent cores at higher frequencies and unchanged ones at lower frequencies. 
This would introduce a steeper gradient of $N(r)$ or $B(r)$ in the region, where the core shift measurement is performed, which would in turn increase the $k_r$, as observed (Table~\ref{jetparam}).}

Taking the measured values of the core shift $\Delta r_{\mathrm{1,2\,mas}}$ and $k_r$ we calculated $\Omega_{r\nu}$ \citep{lobanov_coreshift_1998, mojave_ix_opacity_2012} as follows: 
\begin{equation}
\Omega_{r\nu} = 4.85\cdot10^{-9}\,\, \frac{\Delta r_{\mathrm{1,2\,mas}}\ D_{\mathrm{L}}}{(1+z)^2}\,\, \frac{\nu_1^{1/k_\mathrm{r}}\nu_2^{1/k_\mathrm{r}}}{\nu_2^{1/k_\mathrm{r}}-\nu_1^{1/k_\mathrm{r}}}
\label{omega}
\end{equation}
where $D_\mathrm{L}$ is the luminosity distance.
A deprojected distance between the VLBI core at 7~mm and the jet nozzle was estimated as: 
\begin{equation}
r_{\mathrm{7\,mm}} = \Omega_{r\nu} \left( \nu^{1/k_{r}}\, \sin{\theta} \right)^{-1}.
\label{r_7}
\end{equation}


%
%
%
\begin{table*}
\centering
\begin{tabular}{| l | c | c | c | c |  c | c | c |}
\hline
Epoch &\mb{ \parbox[c]{0.5cm}{\centering   $T_\mathrm{b}^\mathrm{core}$\\($10^{12}$~K)}} & $k_\mathrm{r}$ & \parbox[c]{1.5cm}{\centering $\Omega_{r\nu}$\\\centering(pc GHz)} & \parbox[c]{0.5cm}{\centering $r_{\mathrm{7\,mm}}$\\(pc)}&\parbox[c]{0.5cm}{\centering  $B_1$\\(G)} &\parbox[c]{0.5cm}{\centering  $B^\mathrm{7~mm}_\mathrm{core}$\\(G)} & \parbox[c]{0.5cm}{\centering  $N_1$\\(cm$^{-3}$)} \\ \hline
2009~Aug~28 & \mb{1.1} & $ 0.85 \pm 0.19$ & $ 21.87 \pm 7.66$ & $  2.43 \pm 2.58$ & $  0.47 \pm 0.12$ & $  0.19 \pm 0.21$ & 1500 \\
2009~Oct~25 & \mb{4.7} &$ 0.82 \pm 0.25$ & $ 15.00 \pm 7.68$ & $  1.47 \pm 2.17$ & \mb{$  \mathit{0.34 \pm 0.10}$} & \mb{$  \mathit{0.23 \pm 0.35}$} & \mb{$\mathit{800}$} \\
2009~Dec~05 &\mb{ 6.1} & $ 1.04 \pm 0.29$ & $ 9.63 \pm 3.15$ & $  2.50 \pm 2.64$ & \mb{$  \mathit{0.55 \pm 0.18}$ }& \mb{$  \mathit{0.22 \pm 0.24}$ }& \mb{$\mathit{2100}$} \\
2010~Jan~26 & \mb{5.3} & $ 1.17 \pm 0.27$ & $ 11.42 \pm 2.62$ & $  4.37 \pm 3.40$ & \mb{$  \mathit{0.97 \pm 0.30}$ }& \mb{$  \mathit{0.22 \pm 0.19}$} & \mb{$\mathit{6700}$} \\
\hline
\end{tabular}
\caption{\md{Core parameters. $T_\mathrm{b}^\mathrm{core}$ is the brightness temperature of the 7~mm apparent core. Other parameters are estimated from the core-shift analysis.} $k_\mathrm{r}$ is the power law index; $r_{\mathrm{7mm}}$ -- deprojected distance from the true base of the jet to the apparent core at 7~mm; $B_1$ -- magnetic field strength at a distance of 1~pc along the jet from the true jet base ; $B^\mathrm{7~mm}_\mathrm{core}$ -- magnetic field  at the position of \m{the} apparent core at 7~mm; $N_1$ - particle density at 1~pc from the jet origin. \mb{$B_1$, $B^\mathrm{7~mm}_\mathrm{core}$, and  $N_1$ are estimated assuming equipartition between the magnetic field and particles \mc{energy density}.} \mc{Italic font denotes formally estimated values for epochs with high 7~mm core brightness temperature (Fig.~\ref{tb}) which violates the equipartition  assumption (see Sect.~\ref{britem})}.}
\label{jetparam}
\end{table*}

\ma{The core \citep[$\tau=1$ surface,][]{blandford_koenigl} position depends on both particle density and magnetic field strength \citep[e.g][]{lobanov_coreshift_1998}. \mb{Most of researchers assume equipartition between these two quantities} \citep[e.g.,][]{lobanov_coreshift_1998,hirotani_2005,kutkin_3c454_2014}. 
Even though \cite{zdziarski_2015} \mc{suggested} a method to calculate the jet parameters not assuming equipartition, the accuracy of typical core-shift measurements, including ours, is not sufficient to apply it. Meanwhile, the measured values of the brightness temperature $T_\mathrm{b} > 10^{13}$~K \mc{(Fig.~\ref{tb})} imply that the particles dominate in the total core energy budget. To estimate parameters of the jet base, we first derive the magnetic field and particle density assuming equipartition and then recalculate them for a more plausible set of parameters.}

%


For all four epochs we derive the equipartition magnetic field strength at the distance of 1~pc from the central engine following \cite{kutkin_3c454_2014} as
\begin{equation}
B_1 = 0.014 \left( \frac{\Omega_{r\nu}^{3 k_r} (1+z)^3 \ln({\gamma_{max}}/{\gamma_{min}})}{\phi \delta^2 \sin^{3 k_r - 1}\theta}\right)^{0.25}
\label{B1}
\end{equation}
\ma{where $\gamma_{max}/\gamma_{min}$ ratio is adopted to be $10^3$, the Doppler factor $\delta=6$, the viewing angle $\theta=6^{\circ}$ (see discussion in sect.~\ref{discussion1}) and the corresponding intrinsic opening angle $\phi=1.1^{\circ}$ \citep{pushkarev_jet_opening_angles_2009}.}
To estimate the magnetic field strength \m{at the location of the VLBI core} we have to assume some reasonable values of power law indices $n$ and $m$.
We adopted the equipartition values $n=2$, $m=1$ \citep{blandford_koenigl}. 


\mb{Equipartition implies the following relation between magnetic field and particle density:
\begin{equation}
N_1 = \frac{B_1^2}{8\pi\ln(\gamma_\mathrm{max}/\gamma_\mathrm{min}) m_\mathrm{e} c^2}
\end{equation}
where $m_\mathrm{e}$ is electron mass, and $\gamma_\mathrm{min}$ and $\gamma_\mathrm{max}$ are the electron energy cutoffs.
Equipartition values of $B_1$, $B^\mathrm{7~mm}_\mathrm{core}$, and $N_1$ are presented in Table~\ref{jetparam}.
The values of electron  density at 1 pc are comparable with those presented by \cite{lobanov_coreshift_1998}. However, since we have detected the brightness temperature up to $>10^{13}$~K, the real particle density should be greater than that estimated assuming equipartition. We estimate below the magnetic field strength assuming several \md{increased} values of the particle density.}

\ma{According to \cite{blandford_koenigl} and \cite{lobanov_coreshift_1998} the magnetic field in the 7~mm apparent jet base depends on $\Omega_{r\nu}$ and $N_1$.
We consider the epoch 2009~Aug~28 as a reference one, because physical conditions in the 7~mm core were likely close to equipartition, as deduced from the brightness temperature (Table~\ref{jetparam}), and the estimated B-field should be close to its real value.
We took the measured values of $k_r$ and $\Omega_{r\nu}$ and adopted $\alpha=-0.5$ (hence $k_b=2/3$) and $m=1$ to calculate the ${B'}^\mathrm{7~mm}_\mathrm{core}$ for other three multi-frequency epochs: 
\begin{equation}
{B'}^\mathrm{7~mm}_\mathrm{core} = B^\mathrm{7~mm}_\mathrm{core} \frac{\nu^{1/{k'}_r}}{\nu^{1/{k}_r}} \frac{\big({\Omega'}_{r\nu}/\sin(\theta)\big)^{\frac{{k'}_r}{k_b}-m}}{\big({\Omega}_{r\nu}/\sin(\theta)\big)^{\frac{{k}_r}{k_b}-m}}  \Bigg(\frac{{N'}_1}{{N}_1}\Bigg)^{\frac{-1}{k_b (3/2-\alpha)}},
\end{equation}
\mc{where primed values refer to the epoch of interest and unprimed refer to 2009~Aug~28.}
This formula takes into account both the change of the position of the 7~mm core and the change of the particle density and does not assume equipartition.}
Table~\ref{noneq} presents the magnetic field value at the position of the 7~mm core for different epochs and for 10, 100, and 1000 times increase of the particle density relative to 2009~Aug~28.


\md{We compared our estimates of the apparent 7~mm core B-field with that reported by \cite{savolainen_3c273_2} who have estimated $\log_{10}\Bigl(B^\mathrm{7~mm}_\mathrm{core}\Bigr)=0.3\pm0.6$~G on the basis of synchrotron self-absorbed spectrum turn-over measurements. We estimate the observed brightness temperature of the 7~mm core using parameters provided by \cite{savolainen_3c273_2} to be $T_\mathrm{b} \approx 2\times10^{11}$~K, \m{close} to its equipartition value, \m{relativistically boosted}. Despite the time difference of 6 years and different model-dependent approaches, $B^\mathrm{7~mm}_\mathrm{core}$ of \cite{savolainen_3c273_2} is comparable to that estimated for our epoch 2009~Aug~28, when conditions in the apparent 7~mm core were also close to equipartition. For other three  epochs, exhibiting particle density dominance, according to high $T_\mathrm{b}$, the magnetic field in the 7~mm core should have been lower (Table~\ref{noneq}).}

\begin{table}
\centering
\begin{tabular}{| c | c | c | c |}
\hline
\mb{Epoch} &  \mb{$10\,N$}& \mb{$100\,N$}& \mb{$1000\,N$}\\ \hline
\mb{2009~Oct~25} &  \mb{0.05} & \mb{0.016} & \mb{0.005}\\
\mb{2009~Dec~05} &  \mb{0.08} & \mb{0.025} & \mb{0.008}\\
\mb{2010~Jan~26} &  \mb{0.14} & \mb{0.045} & \mb{0.014}\\
\hline
\end{tabular}
\caption{\mb{Magnetic field at the position of the apparent 7~mm core $B^\mathrm{7~mm}_\mathrm{core}$ (G) for different values of the particle density relative to $N=1500$~$\mathrm{cm}^{-3}$ estimated assuming equipartition for the epoch 2009~Aug~28 (Table~\ref{jetparam}).}}
\label{noneq}
\end{table}

\section{SPECTRAL INDEX analysis}

\begin{figure} 
\includegraphics[angle=0,trim={0.6cm 14.2cm 5cm 1cm},clip]{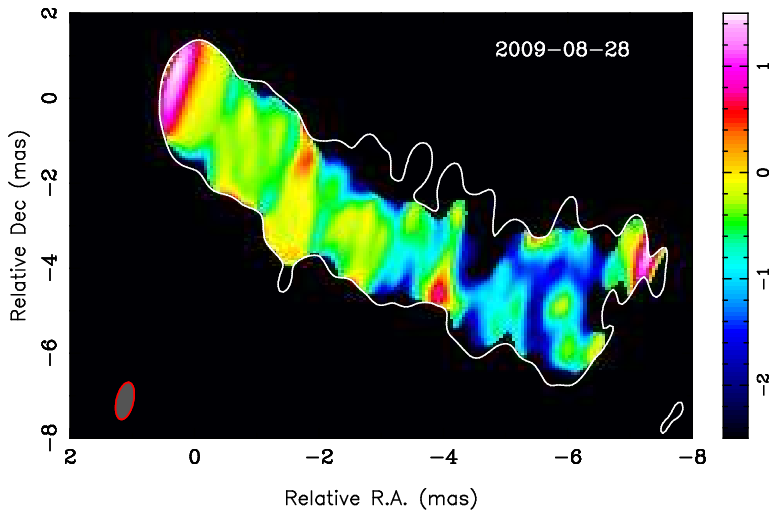}
\includegraphics[angle=270,width=8cm]{{fig10_2}.eps}
\includegraphics[angle=270,width=8cm]{{fig10_3}.eps}
\includegraphics[angle=270,width=8cm]{{fig10_4}.eps}
\caption{Evolution of spectral index between 43~GHz and 24~GHz for all epochs. The contour represents source structure in total intensity at 24~GHz. Contour values per epoch are 9, 8, 13, 22~mJy respectively. 24~GHz beam is plotted in the bottom left corner. Color version for all epochs and frequency pairs is available in online material, Fig.~19-22}
\label{spixmaps}
\end{figure}

After aligning maps we analyzed the two-point spectral index distribution (see Fig.~\ref{spixmaps}) and its evolution during the flaring period. \m{Spectral-index maps were prepared in the same manner as for the core-shift analysis (see Sect.~\ref{coreshift}) with a $5 \sigma$ cutoff applied to the total-intensity maps.}
Examples of spectral-index maps made between 43~GHz and 24~GHz are shown in Fig.~\ref{spixmaps}.

Following \cite{mojave_xi_spectral}, we performed tests to check the impact of different \textit{uv}-coverage (visibility space coverage) on the spectral index maps. For each frequency pair we
produced artificial data in the following manner. We took a model of brightness distribution (result of the  \textit{CLEAN} algorithm) and simulated data on the \textit{uv}-coverages of both frequencies using the \textit{AIPS} task \textit{UVMOD}.  The resulting data were then imaged with parameters identical to those used for real data. Since the brightness distribution the same at both frequencies, spectral index map based on these simulated data should be zero and any discrepancies are due to uneven \textit{uv}-coverage. 
This was done separately for high- and low-frequency \textit{CLEAN} models.
We considered a pixel in the spectral-index map as affected by \textit{uv}-coverage effects if for any of the input CLEAN models the resulting spectral-index value is not zero within the errors. Deviation of this artificial spectral index value from zero tends to increase with distance from the image center. 
According to our tests \textit{uv}-coverage-related effects could be neglected up to certain distances from the map center along the jet. 
For 43~GHz--24~GHz this distance is $4$~mas--$5$~mas, depending on epoch, with one exception, epoch~2010~Jan~26, which was affected strongly because there were only 9 antennas and the \textit{uv}-coverage hence was more sparse. For 24~GHz--15~GHz it is $5$~mas--$6$~mas, for 
15~GHz--8~GHz  -- up to $25$~mas, for 8~GHz--5~GHz -- up to $30$~mas.

\subsection{Core spectral index}
Measuring core spectral index using the flux density of model components at different frequencies might be misleading because we believe the cores at different frequencies are physically different regions in the jet due to opacity. We used a conservative way to estimate the spectral index at the core position. On a two-frequency spectral-index map we derived spectral index value as an averaged value of the 3x3~pixels area around the center of the VLBI core at a lower frequency. 

\begin{figure}
\includegraphics[scale=1]{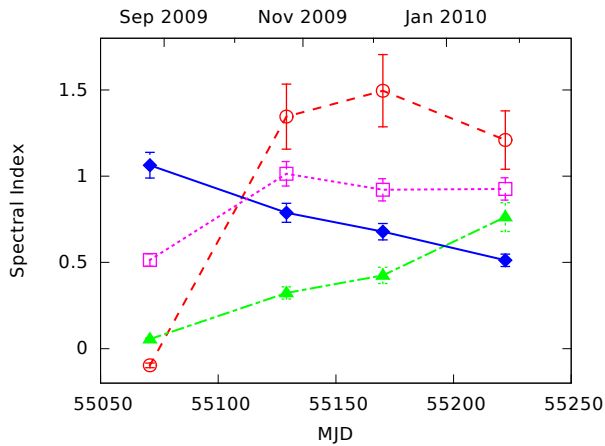}
\caption{Evolution of spectral index $\alpha$ ($S_{\nu}\propto\nu^{\alpha}$) with time in the core at different frequencies. Open circles -- 43~GHz to 24~GHz at position of 24~GHz VLBI core, triangles  -- 24~GHz to 15~GHz at 15~GHz core, diamonds -- 15~GHz to 8~GHz at 8~GHz core, open squares -- 8~GHz to 5~GHz at 5~GHz core.}
\label{corespix}
\end{figure}

We studied the evolution of spectral index between adjacent frequencies during the flare at the apparent jet base. Fig.~\ref{corespix} shows that during all stages of the flare $B_7$, the spectrum in the core at all frequencies was flat or inverted.As expected, higher frequencies respond to the changing physical conditions more swiftly than lower ones.
The spectral index between 43~GHz and 24~GHz exhibits the most remarkable rise from $\alpha_\mathrm{43-24} \approx 0$ to $\alpha_\mathrm{43-24} \approx 1.5$ following the rapidly rising flux density of the 7~mm core. The flux density of the 1.3~cm core also rose during the flare and was reflected in the 24~GHz to 15~GHz spectral index, but not so quickly and with some delay. The spectral index at lower frequencies was almost unchanged during the period considered. This may also indicate that a flare in the cm-band apparent jet base is due to a disturbance carrying larger $N_\mathrm{e}$ that propagates along the jet, passing regions with $\tau \approx 1$ at subsequently lower frequencies. Our measurements did not cover the rise of spectral indexes $\alpha_{\mathrm{15-8}}$ and $\alpha_\mathrm{8-5}$ associated with the flare.

\subsection{Ridge-line spectral index}

\begin{figure}
\includegraphics[width=8cm]{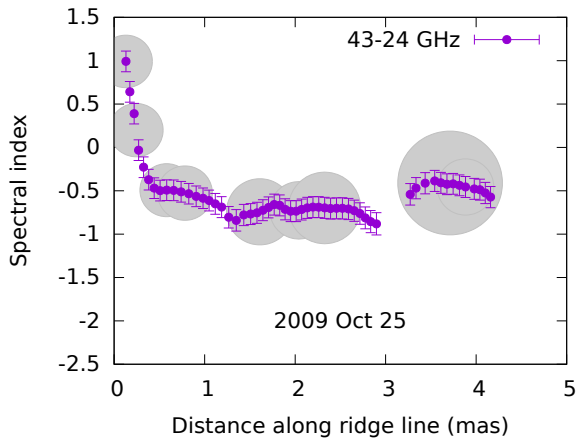}
\caption{43~GHz--24~GHz spectral index values along the ridge line for epoch 2009~Oct~25 with \textit{uv}-coverage effects taken into account. Shaded circles denote the position and size of model components of 24~GHz model convolved with the 24~GHz beam. A full set of ridge line spectral index plots is available in online-only version (Fig.~25-28).}
\label{ridge}
\end{figure}

We analyzed the distribution of spectral index along the ridge line of the jet. \m{For each two-point spectral index map, the} ridge line was calculated at a lower frequency, starting at the position of the VLBI core in a manner similar to that presented by \cite{mojave_xi_spectral}. A plot of the spectral index $\alpha_\mathrm{43-24}$ vs distance along the ridge line for 2009~Oct~25 is shown in Fig.~\ref{ridge}. 
Locations where the ridge line crosses a component are overplotted as a shaded area to compare the component's convolved size with the typical scale of spectral-index variations.

Average values of spectral index along the ridge line are presented in Table~\ref{ridgeavg}. We restricted the range of distances to calculate the spectral index values to the following: $0.6$~mas to $5$~mas for $\alpha_\mathrm{43-24}$, $0.8$~mas to $6$~mas for  $\alpha_\mathrm{24-15}$, $1$~mas to $25$~mas for $\alpha_\mathrm{15-8}$, and $1.7$~mas to $30$~mas for $\alpha_\mathrm{8-5}$. The lower bound is set by the edge of the convolved core (and stationary features for 43~GHz), while the higher bound is set by \textit{uv}-coverage effects.

\begin{table}
\centering
\begin{tabular}{ | c | c | c | c | }
\hline
$\alpha_\mathrm{43-24}$ & $\alpha_\mathrm{24-15}$ & $\alpha_\mathrm{15-8}$ & $\alpha_\mathrm{8-5}$ \\ 
\hline
 $ -0.73 \pm 0.06$ & $-0.81 \pm 0.04$ & $-0.81 \pm 0.07
$ & $-0.68 \pm 0.05
$ \\
\hline
\end{tabular}
\caption{Average value of the spectral-index distribution taken along the ridge line for different frequency pairs.
}
\label{ridgeavg}
\end{table}

Because of our model-fitting strategy (many model components) and restricted distances along the ridge line, according to our tests, ridge lines for all frequencies are almost entirely covered by components. This prevents us from detecting significant deviations of spectral-index values at the positions of components and between them.

We detect a decrease of spectral index, $\Delta\alpha$, with respect to the distance along the jet for $\alpha_\mathrm{15-8}$ and $\alpha_\mathrm{8-5}$ spectral indices on the common range of ridge-line distances $1.7$~mas--$25$~mas. Since \textit{uv}-coverage effects are taken into account we simply fit the spectral index along the jet with a linear \m{function} over a range of distances mentioned above. The average slope of the spectral index along the ridge line is $0.036$~$\mathrm{mas}^{-1}$, which corresponds to decrease of  $10^{-3}$ per deprojected parsec. The value of slope does not differ significantly between  $\alpha_\mathrm{15-8}$ and $\alpha_\mathrm{8-5}$.

\subsection{Spectral index gradient in the transition zone 43~GHz--24~GHz}

We checked whether a transition zone between optically thick (\mc{$\tau\sim1$}) and thin (\mc{$\tau\ll1$}) emission in the core region could be resolved with our VLBA observations. We analyzed slices of spectral index (43~GHz--24~GHz and 15~GHz--8.4~GHz) in the direction of the jet. Errors of spectral index values along the slice are taken from the spectral-index map.

\cite{mojave_xi_spectral} report that a transition from inverted spectrum in the optically-thick core to a steep one in the optically-thin jet is smooth not only due to convolution with the beam but is intrinsic to the jet. We took maps of spectral index between 43~GHz and 24~GHz for all epochs and sliced a region of 1~mas around the core position. A slice starts behind the core and extends in the direction of the jet. The distribution of spectral index we observe is a smoothed intrinsic distribution. 

\begin{figure}
\includegraphics[width=8cm]{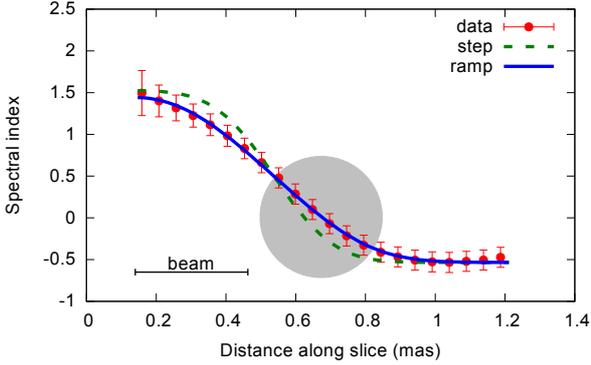}
\caption{Slice oof spectral index between 43~GHz and 24~GHz taken along the jet in transition zone at epoch 2009~Aug~28. Dashed line is a best fit with a \textit{step} function, solid line is a best fit with \textit{ramp} function. \m{Grey shaded circle represents the FWHM and position of the 24~GHz core model component convolved with the beam.} \mc{The \textit{uv}-plane fitted size of the core could be found in online materials, Table~17.} Beam FWHM at \mc{24~GHz} along the slice is shown in lower left corner. Color version for all epochs is available in online material, Fig.~23--24.}
\label{trans}
\end{figure}

There are two ways to estimate the shape of the intrinsic spectral index distribution. First, one may deconvolve the measured distribution to get the intrinsic distribution. Second, one may assume a shape for the intrinsic distribution and convolve it with the beam to compare with the observed distribution. We believe the latter is more appropriate for this kind of study. We considered 2 types of intrinsic distributions: a simple \textit{step} function and a step with linear transition, hereafter \textit{ramp}. We set values of the functions at the ends of the slice to maximum and minimum values of measured spectral index. Thus the \textit{step} function has only a single free parameter, \textit{ramp} has two. We convolved every function with a Gaussian function describing a projection of the beam onto the slice, which resulted in a beam~FWHM~$\approx 0.35$~mas. By varying the free parameters we fitted the convolved function to the data. An example of the best fitted functions for epoch 2009~Aug~28 is presented in Fig.~\ref{trans}. 

The \textit{ramp} function with intrinsically extended transition from $\alpha>0$ to $\alpha<0$ always fits better than a simple \textit{step} function with a sharp intrinsic drop of the spectral index. 
We used corrected Akaike information criterion (AICc) \citep{aic, aicc_book} to test whether \textit{ramp} fits better because it has more free parameters than \textit{step} or does it really better describe the data. The difference in AICcs is very convincing, see Table~\ref{deconv}, favoring the \textit{ramp} function. The resulting relative probability is calculated as $exp((w_\mathrm{step}-w_\mathrm{ramp})/2)$, where $w$ stands for the corresponding AICc value.

\begin{table}
\centering
\begin{tabular}{| c | c | c | }
\hline
Epoch & $w_\mathrm{AICc}$ & \parbox[c]{1.0cm}{\centering Width\\(mas)}\\ \hline
2009~Aug~28  &  $10^5$  &  0.49  \\
2009~Oct~25  &  $10^8$  &  0.45  \\  
2009~Dec~05  &  $18$    &  0.33  \\
\hline
\end{tabular}
\caption{Comparison of models for $\alpha_{43-24}$ spectral index transition zone. $w_\mathrm{AICc}$ shows the relative probability of the best fit \textit{ramp} function in comparison with the best fit \textit{step} function, width (mas) is intrinsic extent of a region with linear descend of spectral index in the fitted  \textit{ramp} function. The last epoch is not included in the table because both models fail to adequately describe the spectral index slice.}
\label{deconv}
\end{table}


We should note that with the given pixel size, adjacent points are not completely independent, which in turn could considerably increase the AICc difference. We have tested a dependence of AICc difference on the number of data points along the slice. We used $\alpha_{43-24}$ and $\alpha_{15-8}$ spectral index data for the epoch 2009~Aug~28. We constructed regular subsamples of the dataset taking every first, every second, every third point etc. 
A relative probability of 0.01 in $\alpha_{43-24}$ is reached when the number of data points is reduced to values small as 10, which corresponds to distance 0.1~mas (1/3 of a beam) between points. 
For the $\alpha_{15-8}$ spectral index the \textit{ramp} model is overwhelmingly more probable regardless of the number of data points. Therefore in our analysis we did not use the whole dataset, but a subsample selected with the intent to leave 5-7 data point per beam FWHM. This resulted in 22 points in 43~GHz--24~GHz and 25 points in 15~GHz--8~GHz .

\begin{figure}
\begin{center}
\includegraphics[width=8cm]{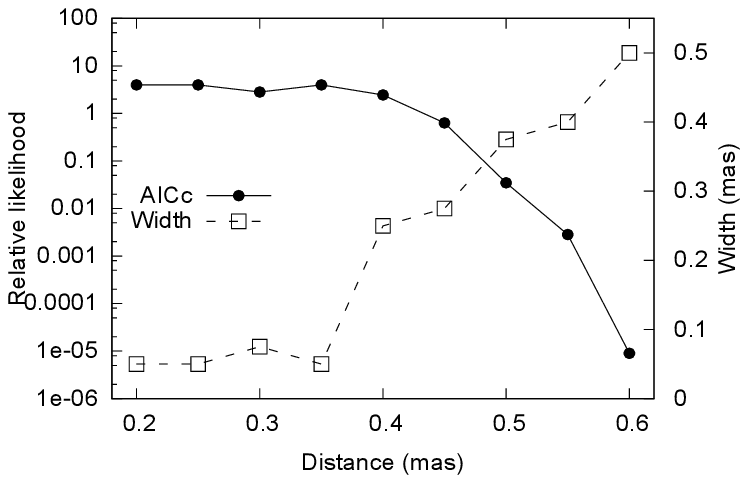}\\
\includegraphics[width=8cm]{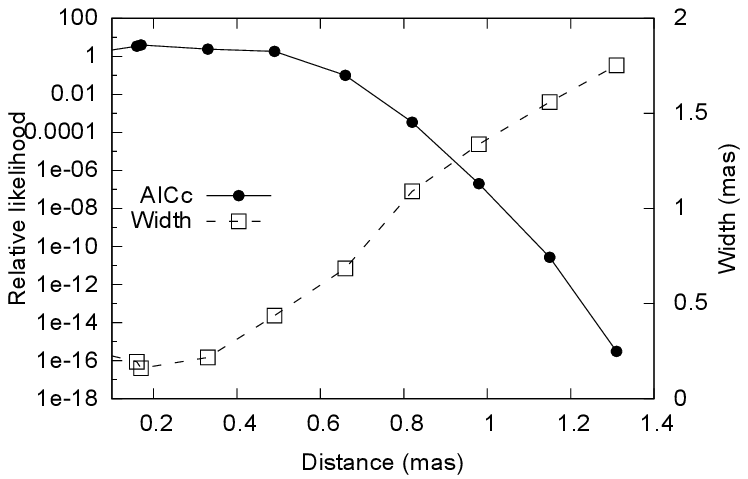}
\end{center}
\caption{Formal probability of the \textit{step} function in comparison to the \textit{ramp} function and width of the transition zone vs distance of slice taken from upstream the  core. 0 is the core center. 43~GHz--24~GHz  (top), every 4th point is plotted. 15~GHz--8~GHz  (bottom), every 12th point is plotted. Epoch 2009~Aug~28.}
\label{aicwidth}
\end{figure}

Another test was performed to check the dependency of the relative model probability on the number of data points removed from beginning of the slice, i.e. from the region upstream the core. This dependence could affect the result if a significant amount of data is removed. 
As shown in Fig.~\ref{aicwidth}, the dependence is not substantial for $\alpha_{15-8}$ until 0.7~mas (0.6 of a beam FWHM) is deleted, which would place a starting point almost onto the upstream core edge. For $\alpha_{43-24}$ the difference between models vanishes when 0.15~mas (0.45 of a beam FWHM) is deleted. Dependency of the determined width of the transition zone on the amount of data taken from behind the core shows a roughly linear trend in data range when \textit{step} and \textit{ramp} models became significantly different.

The minimal estimate of the width of the transition zone for $\alpha_\mathrm{43-24}$ is $0.3$~mas which corresponds to 1~pc projected distance and 7~pc deprojected distance. For $\alpha_\mathrm{43-24}$ most of the transition zone covers a region between the core and the stationary feature \textit{ST2}.

For $\alpha_\mathrm{15-8}$ the estimated width of the transition zone is $2$~mas, and a significant part of the transition zone is located upstream from the 15~GHz core at all epochs.

\section{DISCUSSION}

\subsection{Location of the $\gamma$-ray emission zone}
\label{discussion1}

\m{We considered two effects that might result in a time lag between $\gamma$-ray and radio light curves. First, if both $\gamma$-ray and radio emission is caused by the same perturbation moving downstream from the jet beginning and if a $\gamma$-ray flare is located in a region that is opaque to radio waves then the radio emission is absorbed until the perturbation reaches the partially optically-thin apparent core. 
Second, even if $\gamma$-rays and radio are ignited simultaneously in the same location, transparent for both radiowaves and $\gamma$-rays, then a delay measured between peaks of the flares might be a consequence of a larger size of the radio-emitting region compared to the that of $\gamma$-rays. 
}

Adopting the former scenario and given the delay between light curves from cross-correlation one may calculate a distance between $\gamma$-ray and radio emitting regions as
\begin{equation}
\Delta r = \frac{\delta \Gamma \beta c \Delta t_{\mathrm{7~mm} - \gamma}^{\mathrm{obs}}}{1+z}  = \frac{\beta_{\mathrm{app}} c \Delta t_{\mathrm{7~mm} - \gamma}^\mathrm{obs}}{(1+z) \sin\theta}
\label{deprojdist}
\end{equation}
following \cite{pushkarev_radio2gamma_delay_2010} and the assumptions discussed by them. 


%


The main uncertainty of the $\Delta r$ arises from the poor knowledge of the viewing angle of the jet. Published estimates range from 
$\theta_\mathrm{var}=10\mathring{.}1$ \citep{lahteenmaki_1999} 
to $\theta_\mathrm{var}=3\mathring{.}3$ \citep{hovatta_2009} and $\theta_\mathrm{VLBI} = 6\mathring{.}1\pm0\mathring{.}8$ \cite{jorstad_15agn_2005} and $\theta_\mathrm{VLBI}=9\mathring{.}8\pm2\mathring{.}1$ \citep{savolainen_3c273_1}. According to our $\beta_\mathrm{app}$ estimates on the basis of core and stationary features variability (Sect.~\ref{corest12}) we could derive the maximum value of the viewing angle for $\beta_\mathrm{app}=12.3\,c$: $\theta_\mathrm{max} = \arcsin(2\beta_\mathrm{app}/(1+\beta_\mathrm{app}^2)) = 9\mathring{.}3$ which in turn gives a lower limit on $\Delta r_\mathrm{min} = 6.5$~pc. Estimates of $\theta_\mathrm{max}$ based on component speed measurements, $\beta_\mathrm{app}=8.4\,c$, give $\theta_\mathrm{max} = 13\mathring{.}6$ and $\Delta r_\mathrm{min} = 3.0$~pc.
\mb{We believe the measurements of \cite{jorstad_15agn_2005} are most robust given their excellent temporal coverage, hence, we adopt their $\theta = 6\mathring{}$.}
\ma{With the measured delay of $112$~days and mentioned above range of the jet parameters we estimate the deprojected distance between $\gamma$-ray emitting zone and 7~mm VLBI core to be \m{$\Delta r = 4 - 12$~pc}.}  \mb{For adopted $\theta = 6\mathring{}$ the corresponding $\Delta r = 6.5$~pc is taken as an estimate of the distance between the $\gamma$-ray emitting region and the 7~mm core.} Since $\gamma$-rays lead radio, the $\gamma$-ray emission zone is located closer to the central engine.

The latter scenario implies that both $\gamma$-ray and radio flares start simultaneously. There might be the following associations: $Y_{\gamma}$ starts simultaneously with $Y_{7}$, and both $A_{\mathrm{\gamma}}$ and $B_{\mathrm{\gamma}}$ are consistent with starting contemporaneously with $B_{\mathrm{7}}$ or with the second flare in ST1, and $D_{\mathrm{\gamma}}$ coincides with the start of the second flare in ST2, see Fig.~\ref{corestat}.  

\m{To check if the measured time lag between flares $B_{\gamma}$ and $B_7$ could be fully explained in the latter scenario}, we have compared characteristic rise time of the flare $B_{\mathrm{7}}$ with the light travel time across the core extent. The 7~mm core was unresolved in most observations during the flare $B_{\mathrm{7}}$, so its size was adopted from the resolution limit, which resulted in an average Gaussian FWHM of $a=0.035 \pm 0.005$~mas during the flare. For further calculation we used the core size of $1.6a$ in case the core is a uniform optically-thick disk.  
The expected peak-to-peak delay \m{due to the larger size of the emitting region at 7~mm } in the observer's frame is $\Delta t_{\mathrm{obs}}=208\ \delta^{-1}$~days, \m{where $\delta$ is a Doppler factor}. For $\delta = 4-9$ \citep{lahteenmaki_1999,jorstad_15agn_2005,savolainen_3c273_1} this implies $\Delta t_{\mathrm{obs}}=52$~days--$23$~days and for $\delta = D_{\mathrm{var}} = 17$ \citep{hovatta_2009} $\Delta t_{\mathrm{obs}} = 12$~days. \ma{For adopted  $\delta = 6$, $\Delta t_{\mathrm{obs}} = 34$~days.}

We fitted 7~mm flares $Y_{\mathrm{7}}$ and $B_{\mathrm{7}}$ with exponentials, Eq.~\ref{exp_flares}. For the flare $B_{\mathrm{7}}$ this results in $\tau_{\uparrow} = 82$~days, $\tau_{\downarrow} = 94$~days.  \m{Comparison of $\tau_{\uparrow}$, $\tau_{\downarrow}$  with the expected $\Delta t_{\mathrm{obs}}$ shows that} none of the reported values of the Doppler factor could fully reconcile the light-travel time across the 7~mm core with the variability time scale of the flare $B_{\mathrm{7}}$. Hence we can not treat this effect as the sole factor determining the peak-to-peak lag between 7~mm and $\gamma$-rays, \m{and thus the $\gamma$-ray emitting region should be located upstream from the 7~mm VLBI core.}

Fine structure of the $\gamma$-ray flare in 3C~273 \citep{fermi_3c273, rani_gammasite_3c273_2013} reveals $\gamma$-ray flux variations on the time scale of less than 1 day, which impose certain constraints on the size of the emitting region. Assuming that the $\gamma$-ray emitting region spans the whole width of a conical jet, the variations imply that the distance between the jet nozzle and $\gamma$-ray-emitting region is less than $10^{-3}$~pc and thus is also placed close to the true base of the jet. 



It is worth noting that during the flare \ma{$B_7$} the 7~mm core moved downstream the jet. Since the peak of the flare $B_\mathrm{7}$ occurred when the 7~mm core was at the most downstream position, the cross-correlation analysis gave the maximal value of $\Delta r$. 
If the core-shuttle amplitude of $4.4$~pc is taken into account, the distance from the $\gamma$-ray emitting region to the 7~mm core calculated using the light-curve lag and derived from the core-shuttle analysis is $2.1$~pc to $6.5$~pc, depending on the observational epoch. This agrees well within errors with the distance from the 7~mm VLBI core to the jet nozzle, $1.5$~pc to $4.4$~pc, derived from the core-shift analysis.
This leads us to the conclusion that the $\gamma$-ray emitting region in 3C~273 is located \m{$2-7$}~pc upstream from the 7~mm apparent core, close to the true base of the jet.

\subsection{Passage of moving components through the core, \textit{ST1}, \textit{ST2} and relation to $\gamma$-ray flares}

\begin{figure}
\centering
\includegraphics[width=8cm]{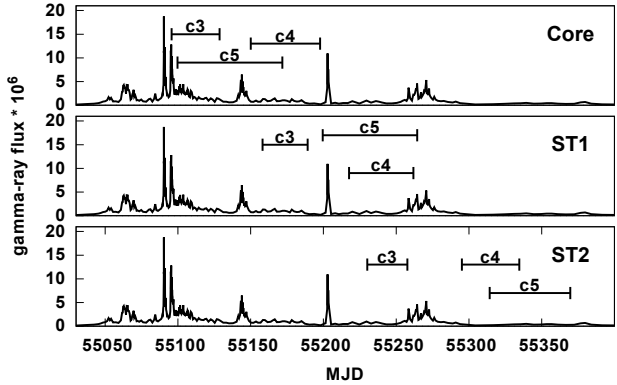}
\caption{Epochs of passing the 7~mm core (upper panel), ST1 (middle), ST2 (lower) for components \textit{c3}-\textit{c5} over the $\gamma$-ray light curve. Uncertainties are adopted from a linear fit.}
\label{crossing}
\end{figure}

Since the jet structure of 3C~273 contains 3 stationary features Core, ST1, and ST2, which could be physical structures like standing shocks, we have performed a cross-identification of all events related to these components with the $\gamma$-ray flares. We calculated epochs when components \textit{c3}--\textit{c5} passed through stationary features. The choice of these components is dictated by their  appearance in the region of stationary features contemporaneously with the period of high $\gamma$-ray activity. In total there were 9 events that might be examined, see Fig.~\ref{crossing}. 
Except for the coincidence of \textit{c5} ejection with $C_{\gamma}$ there is no robust evidence of any correspondence between high-energy flares and transit of components through standing features. 

If, in contrast, a $\gamma$-ray flare was caused by a passing component in the vicinity of the jet apex, then solving Eq.~\ref{deprojdist} for $\Delta t_{\mathrm{7~mm} - \gamma}^\mathrm{obs}$ could link ejected components with $\gamma$-ray flares. This requires an assumption of a constant component speed from the beginning of the jet to the 7~mm core. With \m{the measured $\Delta r = 2-7$~pc and adopted $\theta=6^\circ$}, \textit{c1} and \textit{c2} could have produced $Y_{\gamma}$, \textit{c4} -- either $A_{\gamma}$ or $B_{\gamma}$, \textit{c5} -- $Z_{\gamma}$, and \textit{c3} is left with no $\gamma$-ray counterpart.

The derived ejection epochs for components \textit{c1} and \textit{c2} allow one to connect their passage through the core with the 7~mm core flare $Y_7$. The ejection epoch of component \textit{c4} connects it to the 7~mm core flare $B_7$ and therefore component \textit{c4} could be linked to the preceding $\gamma$-ray flare $B_{\gamma}$. The component appeared after our multi-frequency observations so we have checked parameters of the component derived from only 7~mm data. The component speed is similar to values derived for other components in this research or to values published before \citep{savolainen_3c273_1, mojave_vi_kinem, jorstad_15agn_2005}. We found that this component has maximal brightness temperature $T_\mathrm{b} = 10^{12}$~K in the source frame and a maximal peak flux density $S=6.4$~Jy, the highest among all of the new components ejected between 2008 and 2011. 

\subsection{Core shuttle}
\m{Possible movement of the apparent core, inferred from VLBI kinematics studies, was mentioned previously \citep[e.g.][]{kovalev_opacity_2008, mojave_vi_kinem, hodgson_oj287_2016}. A recent dedicated astrometric experiment \citep{core_wandering_2015} has proven that the core movement along the jet is significant in Mrk~421 and occurs after a strong X-ray flare. There are two effects that could lead to the shuttle of the core. First, if the core position is constant and a new component emerges from the core but is still unresolved from it, one could observe that the core ``moves'' downstream until the new feature becomes resolved from it and then the core jumps back. Second, if physical conditions in the jet change, the core should shuttle down- and upstream. 
This is valid for both interpretations of the mm-core: a $\tau\approx1$ \citep[e.g.][]{lobanov_coreshift_1998} surface or a standing shock \citep{daly_marscher_1988}. }

\m{Using the kinematics of the two clusters of components, we found clear evidence that the apparent core of 3C~273 at 7~mm moved downstream during the onset of the powerful flare $B_\mathrm{7}$ by $4.4$~pc. The core position was anti-correlated with the core flux density. The Pearson correlation coefficient is $r=-0.57$ with a chance probability $p= 10^{-3}$ for the cluster \textit{K1}, and $r=-0.51$, $p=4\times 10^{-3}$ for \textit{K2}. To check if the core shuttle was induced by the flare we performed an additional test by considering independently data during the flare $B_\mathrm{7}$ (MJD 55059-55361) and outside the flare (MJD 55003-55038, 55409-55763). For both clusters there is no significant correlation left for the data outside the flare. Meanwhile during the flare $B_7$ the negative correlation is even more significant. The correlation parameters are summarized in Table~\ref{crco}.}

\begin{table}
\centering
\begin{tabular}{*7c}
\hline
Cluster &  \multicolumn{2}{c}{Flare} & \multicolumn{2}{c}{Non-flare} & \multicolumn{2}{c}{All}\\
{}   & $r$ & $p$ & $r$ & $p$ & $r$ & $p$\\
\hline
\textit{K1}   & $-0.9$ &  $2\times10^{-4}$   & $-0.12$  & $0.36$ &$-0.57$ & $10^{-3}$\\
\textit{K2}   & $-0.93$ & $10^{-5}$  & $-0.13$  & $0.31$ & $-0.51$& $4\times10^{-3}$ \\
\hline
\end{tabular}
\caption{Pearson correlation coefficient $r$ and the permutation test $p-$value for clusters \textit{K1} and \textit{K2} for the period of the 7~mm core flare $B_7$; for the period outside the flare; and for all the data together.}
\label{crco}
\end{table}

\m{A possible explanation of such a prominent negative correlation during the flare is that the core as a $\tau \approx 1$ surface moves along the jet while physical conditions change. A flare in the core could be caused for example by an increased particle density $N_\mathrm{e}$ or increased magnetic field $B$ which also affect the opacity. Higher values of $N_\mathrm{e}$ or $B$ act so that $\tau=1$ occurs farther downstream.}
\mc{The core shuttle together with the frequency dependent core shift \citep{kovalev_gaia_2017} could affect the accuracy of the derived offset between radio VLBI and optical \textit{Gaia} reference frames.}

\section{Conclusions}


\begin{enumerate}

\item Correlated variability of radio and $\gamma$-ray emission, kinematics of a new emerged component and short time scale of $\gamma$-ray variability together with core-shift measurement lead us to the conclusion that the $\gamma$-ray emission site of the most powerful flare $B_{\gamma}$ is located in the nearest vicinity of the true base of the jet in 3C~273, $2$~pc to $7$~pc upstream from the 7~mm VLBI core. This strongest $\gamma$-ray flare might be associated with the brightest moving component \textit{c4}.

\item Kinematic analysis shows that the 7~mm core shuttles along the jet during the 7~mm core flare $B_7$. Core flux density anti-correlates with the core position $r_7$. The total range of the core displacement is $0.17$~mas, or $4.4$~pc along the jet.

\item 
\m{Brightness temperature in the most compact and prominent components at 7~mm, the core and ST1, rises up to $>10^{13}$~K in the source frame. This agrees with previous discovery of the extremely high brightness temperatures in 3C~273 with RadioAstron \citep{kovalev_3c273_2016,johnson_3c273_2016}. Given the moderate measured speed of the model components, this favors the non-equipartition scenario, with the particle energy density dominance.}


\item\md{Magnetic field strength in the 7~mm core is estimated to be $0.2$~G from the core-shift analysis for the epoch 2009~Aug~28. Equipartition between the magnetic field and particles energy density was assumed for this epoch based on the brightness temperature measurements. This magnetic field value is comparable to estimates of \cite{savolainen_3c273_2} made on the basis of synchrotron self-absorbed spectrum turn-over measurements. Other three epochs (2009~Oct~25, 2009~Dec~05 and 2010~Jan~26) were observed by us during a 7~mm flare for which we believe the magnetic field in the core to drop significantly.}

\item{\mc{We estimated the jet bulk speed on the basis of 7~mm light curves of three adjacent stationary jet features. Estimated speed during the flare $B_\mathrm{7}$ was $12c$. This is $1.5$ times higher than the speed of the moving component \textit{c4}, which was associated with the flare, as measured downstream the jet.}}

\item We detected both temporal variations of spectral index in the core and spatial variations along the jet ridge line.
The spectrum of the apparent core at any of the observed frequencies is flat or inverted and becomes harder at high frequencies while a radio flare develops reaching values of $\alpha_\mathrm{43-24}=1.5$ between 43~GHz and 24~GHz. The spectral index steepens along the jet ridge line by $0.036$~$\mathrm{mas}^{-1}$, or $10^{-3}$ per deprojected parsec.

Study of the transition zone from the opaque core to the optically-thin jet at high frequencies shows that this transition extends by several parsecs and could be resolved with our observations at both 24~GHz and 8~GHz.

\end{enumerate}




\section*{Acknowledgments}
\m{We thank the anonymous referee and Alan Roy for thorough reading of the manuscript and constructive suggestions which helped to significantly improve the paper.}
Authors thank Sebastian Kiehlmann for participation in discussion of light curve cross-correlation, and Andrei Lobanov for review and useful discussion. 
Core-shift analysis is supported by Russian Science Foundation grant 16-12-10481. 
Light curve and radio-to-$\gamma$-ray relation analysis is supported by the Russian Foundation for Basic Research (RFBR) grant 13-02-12103.
TH was supported by the Academy of Finland project number 267324.
\mc{TS was funded by the Academy of Finland projects 274477 and 284495.}
This research has made use of data from the MOJAVE database that is maintained by the MOJAVE team 
\citep{mojave_vi_kinem}. This study makes use of 43~GHz~VLBA data from the VLBA-BU Blazar Monitoring Program (VLBA-BU-BLAZAR; http://www.bu.edu/blazars/VLBAproject.html), funded by NASA through the Fermi Guest Investigator Program. The VLBA is an instrument of the National Radio Astronomy Observatory. The National Radio Astronomy Observatory is a facility of the National Science Foundation operated by Associated Universities, Inc.
This research has made use of data from the University of Michigan Radio Astronomy Observatory which has been supported by the University of Michigan and by a series of grants from the National Science Foundation, most recently AST-0607523.

\bibliographystyle{mnras}
\bibliography{lisakov_radio2gamma}

\appendix

\bsp	
\label{lastpage}

\end{document}